\documentclass[manuscript, natbib=false]{acmart}
\usepackage[utf8]{inputenc}
\usepackage{amsmath}
\usepackage{amsthm}
\usepackage{amssymb}
\usepackage{amsfonts}
\usepackage{xcolor}
\usepackage{appendix}
\usepackage{graphicx}
\usepackage{url}
\usepackage{subfiles}
\usepackage{caption}
\usepackage{subcaption}
\usepackage{enumitem}
\usepackage{hyperref}

\renewcommand\footnotetextcopyrightpermission[1]{} 
\setcopyright{none}
\usepackage{xpatch}

\makeatletter
\xpatchcmd{\ps@firstpagestyle}{Manuscript submitted to ACM}{}{\typeout{First patch succeeded}}{\typeout{first patch failed}}
\xpatchcmd{\ps@standardpagestyle}{Manuscript submitted to ACM}{}{\typeout{Second patch succeeded}}{\typeout{Second patch failed}}    \@ACM@manuscriptfalse
\makeatother

\newtheorem{mech}{Mechanism}

\newcommand{\norm}[1]{\left\lVert#1\right\rVert}

\begin{document}

\title{Multi-Category Fairness in Sponsored Search Auctions}

\author{Shuchi Chawla}
\affiliation{University of Wisconsin-Madison}
\authornote{shuchi@cs.wisc.edu, supported in part by NSF grants CCF-1617505 and CCF-1704117.}

\author{Christina Ilvento}
\affiliation{Harvard University}
\authornote{cilvento@g.harvard.edu, funded in part by the Sloan Foundation.}

\author{Meena Jagadeesan}
\affiliation{Harvard University}
\authornote{mjagadeesan@college.harvard.edu}

\begin{abstract}
Fairness in advertising is a topic of particular concern motivated by theoretical and empirical observations in both the computer science and economics literature. We examine the problem of fairness in advertising for general purpose platforms that service advertisers from many different categories. First, we propose inter-category and intra-category fairness desiderata that take inspiration from individual fairness and envy-freeness. Second, we investigate the ``platform utility'' (a proxy for the quality of the allocation) achievable by mechanisms satisfying these desiderata. More specifically, we compare the utility of fair mechanisms against the unfair optimal, and we show by construction that our fairness desiderata are compatible with utility. That is, we construct a family of fair mechanisms with high utility that perform close to optimally within a class of fair mechanisms. Our mechanisms also enjoy nice implementation properties including metric-obliviousness, which allows the platform to produce fair allocations without needing to know the specifics of the fairness requirements. 

\end{abstract}

\maketitle

\section{Introduction}\label{sec:introduction}
In the ongoing discussion of what it means for automated decision-making systems to be fair, the topic of online advertising has merited particular interest.
In the United States, segregated employment ads for men and women proved to be a flashpoint in the 1960s, and the introduction of ever-more finely-tuned advertising online has renewed concerns about discrimination in ads for critical categories such as employment, housing and credit.
Although individual advertisers certainly have opportunities to use fine-grained targeting (in some cases, targeting individual users) to implement biased advertising strategies, there is both empirical evidence and theoretical support for the idea that the troubling trends in skewed advertisement between different demographic groups do not occur solely because of bad actors among advertisers.

In fact, recent work has shown that {\bf even when the advertisers all act fairly in isolation, revenue-optimized platform mechanisms can result in unfairness.}
In particular, competition between advertisers in first-price auctions, particularly across categories, can introduce a significant skew in the types of ads people see (e.g., \cite{LT2016}). For example, two equally qualified software engineers may see wildly different numbers of employment ads depending on \textit{competition} for their attention from other categories.
Competition from lucrative categories like children's products can be difficult for an individual advertiser to correct for, particularly in categories like employment where parental status is considered very sensitive.

While the idea of parents being excluded from employment ads due to competition is inherently troubling, blunt policies like equalizing the number of ads across users within {\em each} category can also result in poor outcomes for individuals. 
For example, suppose Bob is highly qualified for a new credit card, and he is also searching for daycare options in his area. Alice is equally qualified for the credit card, but she is not interested in daycare. Bob would not want to see fewer ads for daycare services to ensure that he sees as many ads for credit as Alice, and likewise Alice would not want good credit offers suppressed in order to ensure that she doesn't see more credit ads than Bob who doesn't even care about credit cards anyway!

Competition across ads \textit{within} a single category can likewise result in skew. Even if two equally qualified individuals see the same number of ads in a category, these ads may not be equally relevant or valuable to them. For example, these individuals could see the same total number of job ads, but with one receiving ads for jobs with significantly higher salaries than the other.\footnote{ Subtle issues, like willingness to relocate, company size and type, and other policies may also significantly influence the usefulness of the ads to each user.}

How can we formalize these intuitive desiderata? 
Are these fairness desiderata compatible with consumer, platform, and advertiser utility?

\subsection*{Our Contributions}
\subsubsection*{Model and Definitions}
In this work, we examine the problem of fairness in advertising across multiple categories from a theoretical perspective. We propose a stylized model and fairness requirements that match the intuitive fairness desiderata of inter- and intra-category guarantees.\footnote{Our aim is to propose a model and examine its properties to gain insight into how certain notions of fairness interact with platform utility as well as our intuition about what is fair rather than to propose a specific notion of fairness for practical implementation.} Moreover, our fairness requirements incorporate the qualifications of users as well as their preferences across different ad categories. Our definitions are inspired by and combine the complementary notions of \textbf{envy-freeness} and \textbf{individual fairness}. Envy-freeness~\cite{caragiannis2012efficiency} requires that every user should prefer their own allocation to that of everyone else; it ignores users' qualifications and considers preferences as paramount. Individual or metric fairness~\cite{DHPRZ2012}, on the other hand, ignores preferences and essentially requires that similar users should be treated similarly. Its extension to multi-dimensional allocations as introduced in \cite{DI2018}, \textbf{multiple task fairness}, requires that individual fairness is satisfied separately and simultaneously for all categories.


For inter-category competition, we note that multiple task fairness is at odds with consumer and platform utility. Consider the example of Alice and Bob discussed above. Taking this example to the extreme suppose that a user, whom we will call Jack, is a ``jack of all trades,'' i.e., qualified or interested in many categories. Per multiple task fairness, another user who is as qualified as Jack in a single category but unqualified in all other categories, would have their allocation within the single category of interest limited to match Jack's allocation within that category. Multiple task fairness would then either (1) enforce a minimum amount of exposure for Jack to ads they may not care about or (2) dictate a maximum amount of exposure to relevant ads for other users who are qualified for a smaller number of categories. In effect, by ignoring users' own preferences, this stringent notion of fairness helps no one. We thus introduce \textbf{inter-category envy-freeness}, which allows users to specify a set of categories that they ``care'' about, and guarantees that they see at least as many ads that they care about as any other individual.\footnote{Kim et al. \cite{KKRY2019} has similar motivation, but takes a different approach to blending user-preferences and individual fairness.}

For intra-category competition, we show that multiple-task fairness can be too weak to avoid discriminatory allocations and is therefore vulnerable to certain subversion attacks. For example, if Alice sees {\em every} high-paying job ad slightly less often than Bob and {\em every} low-paying job ad slightly more often than Bob, then for any single ad the two users obtain similar allocations, but across the set of all high-paying jobs, Alice may receive a far smaller share than Bob. We thus introduce \textbf{total variation fairness}, which offers protection against these subversion attacks and captures fairness over all possible sets of substitutes.\footnote{Although ``high-paying'' or ``low-paying'' are well-defined attributes in the employment setting, there are many subtle properties which may be harder to articulate, e.g., ``commute time.'' By protecting all sets of substitutes, we obviate the need to enumerate all possible attributes.}

Combining these fairness aims yields a definition giving fairness both across categories and within categories. More specifically, we demonstrate how to combine inter-category envy-freeness and total-variation fairness into a hybrid fairness notion that we call \textbf{compositional fairness}. For example, suppose two individuals, Alice and Bob, are equally qualified and interested in a particular type of job. Alice is also interested in the latest movie releases, while Bob is interested in buying a new car. We provide the guarantee that Alice and Bob see the same \textit{mix} of ads for jobs, but are permitted to see job ads overall with different probabilities, so long as each sees their preferred categories in total at least as often as the other.

\subsubsection*{Fair mechanisms achieving high utility}
In practice, fairness is obviously not the only consideration in mechanism design. Platforms typically aim to optimize for (1) short-term revenue, i.e. the sum of payments made by advertisers for displaying their ads, or (2) allocative efficiency a.k.a. social welfare, i.e. the sum of the perceived values of the user-ad matches, or a combination of these objectives. As a proxy for these complex objectives, we use the sum of the bids of the ads displayed as a measure of the quality of an allocation. We call this objective {\bf platform utility}. High platform utility correlates both with advertiser happiness (ads get matched with targeted ``high bid'' users) as well as with platform revenue (advertisers often pay their bid or an increasing function of their bid). While our goal is to design {\em fair} mechanisms that achieve high platform utility, we compare the performance of our mechanisms to the {\em unfair} optimum, following the intuition that platforms and advertisers are unlikely to adopt mechanisms which deviate wildly in utility from the status quo. In this respect, we show that our fairness desiderata are compatible with high platform utility.


Apart from these fairness and utility objectives, from a mechanism design perspective, it is desirable for our mechanisms to satisfy certain implementation properties. For example, the platform may like to have some ability to outsource auditing of fairness instead of taking on the responsibility of determining what constitutes fairness in each category and enforcing those determinations.\footnote{Particularly in cases in which determining fairness requires sensitive information (or sensitive information may be leaked by the determination procedure) the platform is likely to want to avoid collecting or possessing such information.} In this context, the platform would likely prefer to be oblivious to the particulars of fairness requirements, and instead make a guarantee that if each advertiser behaves ``fairly'', then the platform does not introduce any additional unfairness. With this guarantee, the platform would thus be able to leave auditing of advertiser behavior to a neutral third party or governing body.

Our main result is a family of mechanisms that achieve compositional fairness and utility close to optimal within the class of fair mechanisms, while also being oblivious to fairness requirements. The nature of the fairness definitions enables us to compose a mechanism satisfying total-variation fairness within each category with a mechanism satisfying inter-category envy-freeness across categories to obtain a mechanism satisfying compositional fairness. Thus, we can separately design inter-category mechanisms and intra-category mechanisms. The inter-category mechanisms that we propose are always envy-free regardless of the bids, and the intra-category mechanisms that we propose are oblivious to fairness requirements: our intra-category mechanisms implicitly use fairness of advertiser bids to infer the necessary fairness requirements. Unsurprisingly, if advertisers are allowed to bid very differently on similar individuals, it is impossible to simultaneously achieve fairness and high platform utility. As part of our study, we also explore upper and lower bounds on the strength of a bid fairness condition needed to achieve a high platform utility.


\subsection*{Outline for the rest of the paper}
The remainder of this work is organized as follows: related work is discussed in Section \ref{sec:related}; definitions and our formal model are introduced in Section \ref{sec:definitions}; mechanisms achieving intra-category fairness are discussed in Section \ref{sec:identical}, and mechanisms achieving inter-category envy-freeness are discussed in Section \ref{sec:different}; composing these mechanisms is addressed in Section \ref{sec:combo}; finally future work is discussed in Section \ref{sec:future}. We defer proofs and some discussion to the Appendix.

\section{Related Work}\label{sec:related}
Fairness in advertising is a topic of particular interest in the popular press, the empirical and theoretical computer science literature, as well as economics.
There are number of compelling empirical studies and popular press articles demonstrating ``skew'' in advertisement delivery between groups, either due to advertiser targeting choices, platform delivery issues, or competition between advertisers \cite{AP2016, BTI2019, AST2017, LT2016, ASBKMR2019}. 

In the theoretical computer science literature, Dwork et al. \cite{DI2018} proposed the notion of individual fairness as a fairness concept for settings including advertising. Dwork and Ilvento \cite{DI2018} consider individual fairness (and group fairness) in advertising as a composition problem, and poses a limited set of fair composition mechanisms. Similar to this work, Kim et al. \cite{KKRY2019} propose a notion of ``Preference Informed Individual Fairness'' (PIIF) to capture the idea that deviation from individual fairness is acceptable as long as it is based on individuals' preferences. However, while PIIF is intended to expand the definition of fairness within the context of task-specific metrics, our preferences-based envy-freeness concept applies across multiple categories independent of the metric for each category.
Furthermore, whereas Kim et al. focus primarily on optimizing user utility via offline optimization across all fair outcomes, we study online mechanisms and measure utility relative to the unfair optimum. We view our work as complementary to \cite{KKRY2019}, and we anticipate that combining the insights from their perspective and ours will be useful for proposing alternative mechanisms and fairness relaxations.

Other works on fair advertising and related problems employ different notions of fairness. With respect to group or statistical notions of fairness, Mehrotra et al. \cite{celis2019toward} propose a mechanism for ad delivery while maintaining certain group level statistics. A variety of fairness notions have also been considered in related problems such as ranking \cite{celis2017ranking, evidencerankings2019}, recommender systems \cite{antikacioglu2018new}, and news search engines \cite{Kapoor2018}. 

Finally, mostly disjoint from the work referenced above, there is an extensive literature in fair division concerning other notions of fairness \cite{caragiannis2012efficiency}. For one of these fairness notions, called ``envy-freeness'', the goal is to partition a limited shared resource among multiple agents with heterogenous preferences. A major difference between this literature and the fairness literature referenced above is that envy-freeness is defined exclusively based on agents' preferences and not on their traits or qualifications. Two recent works, Balcan et al. \cite{balcan2018envy}  and Zafar et al. \cite{zafar2017parity} seek to combine notions of envy-freeness with parity-based fairness in machine learning in the context of classification.

\section{Advertising and Fairness Models}\label{sec:definitions}

\newcommand{\bfp}{{\bf p}}

\newcommand{\bfd}{{\bf d}}

We model the online advertising problem as follows. A universe $U$ of users arrive in an online fashion. There are $k$ advertisers indexed by $i\in [k]$. When a user $u$ arrives, each advertiser $i$ places a bid $b_u^i \geq 0$ on the user. The allocation algorithm then assigns allocation probabilities $p_u^i \in [0, 1]$ to the advertisers with $\sum_{i=1}^k p_u^i \le 1$. The allocation mechanism is an online algorithm, i.e. it assigns allocation probabilities without observing bids on users that arrive in the future or the ordering of future user arrivals. Moreover, we assume for simplicity that each user arrives at most once. We use $\bfp$ to denote the allocation rule output by the allocation algorithm.

The goal of the allocation mechanism is to maximize the sum of the bids of the ads displayed. Formally, this is given by:
\[\textrm{Utility}(\bfp) = \sum_{u\in U} \sum_{i\in [k]} p_u^i b_u^i.\]
We measure the utility of our mechanisms against the best achievable in the absence of any fairness constraints.
The utility is easy to maximize in the absence of any constraints on how allocations vary across users: the mechanism can simply assign a probability mass of $1$ to the highest bidder for every user.\footnote{This is equivalent to running a first-price auction.} We call the corresponding utility the unfair optimum:
\[\textrm{Unfair-OPT} = \sum_{u\in U} \max_{i\in [k]} b_u^i.\]
The {\bf Fair Value} of an allocation mechanism is the ratio of its utility to the Unfair-OPT. Note that this ratio is always less than 1; the larger the fair value the better the utility of the allocation is.

Why compare with the \textit{unfair} optimal utility instead of the fair optimal (as do other related works \cite{KKRY2019, celis2019toward})? First, the utility of the fair optimal mechanism can be difficult to analyze due to the fairness requirements being revealed in an online fashion. Second, if there is a large gap between the utility of the mechanisms we propose and the utility of the unfair-optimal mechanism, the platform and advertisers may be unwilling to adopt the mechanism, so it is critical that we compare with the platform's status quo.

\subsection{Fairness of allocation}

Our fairness guarantees are based on the concept of individual fairness defined by \cite{DHPRZ2012} and the well-studied concept of envy-freeness\footnote{See \cite{caragiannis2012efficiency}.}. At a high level, individual fairness guarantees that similar individuals are treated similarly. Similarity between individuals is captured through a fairness metric $\bfd$ over $U$ and similarity between outcomes is captured by defining a metric $D$ over distributions over outcomes.
\begin{definition}[Individual Fairness \cite{DHPRZ2012}]
A function $f: U \rightarrow \Delta(O)$ assigning users to distributions over outcomes is said to be \textbf{individually fair} with respect to distance metrics $\bfd$ over $U$ and $D$ over $\Delta(O)$, if for all $u,v\in U$ we have $D(f(u), f(v)) \le \bfd(u,v)$.
\end{definition}

Dwork and Ilvento \cite{DI2018} proposed extending the notion of individual fairness to settings involving multi-dimensional allocations by ensuring fairness separately within each dimension or ``task''. This gives rise to the notion of multiple-task fairness, which we now define in the context of online advertising. Let $\{C_1, \ldots, C_c\}$ denote a partition of the set $[k]$ of advertisers into $c$ categories. For $1 \le j \le c$, let $\bfd^j$ denote a pseudometric over the users relevant to all advertisers in category $j$; $\bfd^j: U \times U \rightarrow [0,1]$. In our setting, the outcome assigned to each user $u$ corresponds to the advertiser who is assigned to the slot for user $u$. Our mechanism maps users to distributions over outcomes, i.e. to the allocation probabilities $\left\{p_u^i \right\}_{1 \le i \le k}$. We use the absolute difference between these probabilities to capture the similarity of allocations, and multiple-task fairness becomes the following condition:\footnote{We can also view the assignment $\left\{p_u^i\right\}_{1 \le i \le k}$ as a fractional allocation. In this case, the distance corresponds to the difference in the portion of allocation for each advertiser.}
\begin{definition}[Multiple-Task Fairness \cite{DI2018}]
An allocation function $\bfp$ satisfies \textbf{multiple-task fairness} with respect to distance metrics $\{\bfd^j\}_{j\in [c]}$ if for all $u,v\in U$, $j\in [c]$, and $i\in C_j$, we have $|p^i_u - p^i_v| \le \bfd^j(u,v)$.
\end{definition}
\noindent We will demonstrate that multiple-task fairness is too weak for fairness within a single category, and it results in suboptimal allocations across categories from the perspective of envy-freeness. We propose two new multi-dimensional fairness notions, one applying across different categories and the other to multiple advertisers within the same category, and combine these into the notion of \textbf{compositional fairness} that overcomes the shortcomings of multiple-task fairness.

\subsubsection*{Intra-category fairness}
First, we consider a setting in which all of the advertisers belong to a single category (i.e. $c = 1$) with a single metric that we denote by $\bfd$ (e.g., a tech job-search website containing advertisements only from tech employers).
We observe that multiple-task fairness is insufficient to protect against two broad classes of problems.
\begin{enumerate}[leftmargin=*]
    \item \textbf{Intentional unfairness:} It is vulnerable to subversion by malicious advertisers.
In particular, consider an advertiser that submits multiple different ads (pretending to be distinct advertisers) for the same job and bids separately on each user for each of those ads. The advertiser effectively poses as multiple sub-advertisers; let $S$ denote the set of these sub-advertisers. In this case, the multiple-task fairness constraint only ensures $|p^i_u - p^i_v| \le \bfd(u,v)$ for each $i$, and so it is possible that  $|\sum_{i \in S} p^i_u - \sum_{i \in S} p^i_v| = |S| \bfd(u,v)$. As a result, the advertisers may be able to amplify the difference in probabilities of allocation between two users to an arbitrarily large extent.
\item \textbf{Unintentional unfairness:} Multiple-task fairness can also interact in undesirable ways with well-intentioned advertisers. Suppose that the set $S$ consists of all high-paying job ads.\footnote{The set $S$ can also be job ads from a certain geographical area.} Suppose high-paying advertisers all bid higher on user $u$ than on user $v$, and the mechanism sets $|p_u^i - p_v^i| = \bfd(u,v)$ for all $i \in S$ (to maximize utility). Then, it would again be the case that $|\sum_{i \in S} p^i_u - \sum_{i \in S} p^i_v| = |S| \bfd(u,v)$, so the total allocation on high-paying job ads (i.e. advertisers in $S$) can be vastly different for $u$ and $v$.
\end{enumerate}

To rectify these issues, we propose \textbf{total variation fairness} which requires that the allocation vectors $p_u$ and $p_v$ are not only close component-wise, but are also close in terms of $\ell_1$ distance or total variation distance.
\begin{definition}[Total Variation Fairness]
An allocation function $\bfp$ satisfies \textbf{total variation fairness} with respect to a metric $\bfd$ if for all $u,v\in U$ and all $S\subseteq [k]$, we have $|\sum_{i \in S} p^i_u - \sum_{i \in S} p^i_v| \le \bfd(u,v)$. Equivalently, for all $u,v\in U$, $\norm{p_u-p_v}_1 \le \bfd(u,v)$.
\end{definition}
\noindent This stronger definition, which provides guarantees on all subsets of advertisers $S$, effectively mitigates the issues outlined above.
First, it prevents the multiple bid attack by ensuring that $|\sum_{i \in S} p^i_u - \sum_{i \in S} p^i_v| \le \bfd(u,v)$. Second, it provides nice guarantees over substitutes in the following sense. Consider a user $u$ who regards some arbitrary subset $S$ of advertisers to be substitutes. In that case, the probability that the user observes an ad from this subset is $\sum_{i\in S} p_u^i$ and total variation fairness ensures that this sum is close to the corresponding sum for similar users.\footnote{This definition can be viewed as combination of the multiple-task fairness and OR-fairness definitions put forth in \cite{DI2018}. The definition essentially provides OR-fairness over all possible subsets of advertisers.}

\subsubsection*{Inter-category fairness} Next consider a setting where every advertiser belongs to a different category, i.e. where $c = k$. As the ``jack of all trades'' example in the introduction (formalized in Section \ref{sec:different}) shows, multiple-task fairness can lead to outcomes that, although technically fair, are undesirable from every stakeholder's perspective -- (1) users get low allocations in their desired categories, (2) advertisers reach far fewer qualified users and (3) the platform gets low utility owing to the poor quality of the matching produced. In this example, allocations available to the jack of all trades are constrained by the limited attention of this single user (a single ad slot in our model). Multiple-task fairness combined with this constraint limits the allocations of all of the other users, thereby hurting their utility, without in turn providing any benefit to the jack of all trades.  Within this context, we view the multiple task fairness objective to be unduly skewed in favor of a single individual over the collective good.


Is it possible to achieve a better balance? It is if we slightly shift our viewpoint. Consider a platform where each user is allowed to choose the category they are most interested in with the guarantee that they see at least as many ads in this category as {\em any other} user.
In effect, no user is envious of other users given their own choice of preferred category. A few remarks are in order. First, in the special case where every category has a single advertiser,\footnote{We discuss the case of multiple advertisers per category and allocation between those advertisers in the subsection on compositional fairness.} within the chosen category, the fairness guarantee provided to the user is much stronger than that guaranteed by individual fairness -- the user does not just obtain an allocation close to that of other similar users, but rather obtains one that is as good or better than that of everyone else.\footnote{The reader may worry that a user who is ``unqualified'' in their chosen category could obtain a large allocation at the expense of advertisers that don't want to target such a user. Note that advertisers can bid $0$ on users that are not targeted and thereby pay nothing for those users. Furthermore, it is rare for a user to be wholly unqualified for a category from the perspective of \textit{advertising}. For example, an individual who is unqualified for a job is likely a good candidate for a job training ad.} Second, this is balanced by the lack of any fairness guarantee on non-chosen categories. However, note that from the user's perspective those other categories are anyway not important.
To take an example, suppose that Alice and Bob are identical in terms of their credit-worthiness as well as job qualifications. Suppose Alice is looking for a job and Bob for a credit card. Then a platform that shows a job ad to Alice and a credit card ad to Bob makes both users happy and envy-free. In contrast, a platform that shows each user one of the two ads uniformly at random is multiple-task fair but makes both users worse off. Third, while our definitions focus on a single arrival of each user, one may envision a system where a user interacting with the platform multiple times can change their preferred category at each interaction and thereby obtain a fair allocation within the chosen category at each individual interaction. Finally, observe that the envy-freeness guarantee is directional and entirely independent of distance metrics. 

More generally, we allow users to select multiple preferred categories and provide an envy-freeness guarantee with respect to the total probability of seeing an ad within their set of preferred categories. Formally, each user $u$ picks a preferred set $S_u\subseteq [c]$ of categories.\footnote{From an implementation perspective, we might imagine that users specify these categories on a user profile.} We then ensure that $\sum_{i\in S_u} p^i_u$ is at least as large as the corresponding sum for any other user $v$.
\begin{definition}[Inter-Category Envy-Freeness]
An allocation function $\bfp$ satisfies \textbf{inter-category envy-freeness} with respect to preferred sets $\{S_u\}_{u\in U}$ if for all $u,v\in U$, we have $\sum_{i \in S_u} p^i_v \le \sum_{i \in S_u} p^i_u$.
\end{definition}

\subsubsection*{Compositional fairness}
Now we consider the general setting where there can be multiple categories and multiple advertisers in each category. We discuss how to combine the two definitions above to provide hybrid fairness guarantees. We have two goals: (1) each user should be envy-free with respect to the categories of ads they see and (2) within each category, the mix of ads presented to each user satisfies our strengthened notion of individual fairness. For example, suppose that two similarly qualified users Alice and Bob both select jobs as their preferred ad category. Not only should the two users then see job ads with the same total likelihood, but they should also see a similar mix of high-paying and low-paying job ads.

This composition of definitions becomes subtle when users select \textit{multiple} preferred categories. Suppose that Alice continues to choose jobs as her preferred category, but Bob chooses both jobs and household product ads. Suppose, further, that Alice is allocated a jobs ad with probability $1$ and Bob sees an ad in each of the two categories with probability $1/2$ each. This allocation satisfies inter-category envy-freeness. However, within the job ads category there is no way to assign probabilities that satisfy unconditional total variation fairness simply because of the fact that we have different total probabilities to distribute. Intuitively, we want Bob to be able to see the same {\em mix} of ads as Alice even though Alice may see job ads more frequently overall. Accordingly, we enforce total variation fairness on the {\em conditional distribution} of allocation within each category.

Formally, we define compositional fairness as follows. We use $\bfd^j$ to denote the metric specific to category $C_j$ and $q^j_u = \sum_{i \in C_j} p^i_u$ to denote the total allocation within category $C_j$ for user $u$.
\begin{definition}[Compositional Fairness]
An allocation function $\bfp$ satisfies \textbf{compositional fairness} with respect to distance metrics $\{\bfd^j\}_{j\in [c]}$ if the assignments $\left\{q^j_u\right\}_{u \in U, j\in [c]}$ satisfy inter-category envy-freeness, and for each $j\in [c]$ such that $q^j_u > 0$, the conditional probabilities $\left\{\frac{p^i_u}{q^j_u} \right\}_{i \in C_j}$ satisfy total variation fairness with respect to $\bfd^j$.
\end{definition}

\subsubsection*{Multiplicative Relaxations}
We can further refine each of the above notions by defining multiplicative relaxations parameterized by $\beta \in [1, \infty)$:
\begin{itemize}[leftmargin=*]
\item {\bf $\beta$ total variation fairness:} for all $u,v\in U$ we have $\norm{p_u - p_v}_1 \le \beta \bfd(u,v)$.
\item {\bf $\beta$ inter-category envy-freeness:} for all $u,v\in U$ we have $\sum_{i \in S_u} p^i_v \le \beta \sum_{i \in S_u} p^i_u$.
\item {\bf $\beta$ compositional fairness:} $\left\{q^j_u\right\}_{u \in U, j\in [c]}$ satisfies $\beta$ inter-category envy-freeness, and for each $j \in [c]$ such that $q^j_u > 0$, $\left\{\frac{p^i_u}{q^j_u} \right\}_{i \in C_j}$ satisfies $\beta$ total variation fairness.
\end{itemize}

\section{Intra-category fairness}\label{sec:identical}
\newcommand{\bfd}{{\bf d}}
 
In this section, we focus on the case where advertisers are in a single category $(c = 1)$, i.e. where advertisers face the same metric over users, in particular, $\bfd = \bfd^1 = \bfd^2 = \ldots = \bfd^k$. 

In Section \ref{subsec:bidratio}, we describe the need for a fairness condition on advertiser bids, that we call a bid ratio condition. In Section \ref{subsec:uniform}, we investigate the special case of uniform metrics and establish impossibility results, i.e. upper bounds on the fair value of any allocation mechanism that satisfies multiple-task fairness as a function of the bid ratio condition. In Section \ref{subsec:gensimilar}, we consider settings with arbitrary distance metrics and exhibit an allocation mechanism that is metric-oblivious, history-oblivious, and achieves total variation fairness with respect to the given metric with an appropriate bid ratio condition. We then bound the fair value achieved by this mechanism as a function of $k$ (the number of advertisers) and a parameter defining the bid ratio constraint. Moreover, we show that this mechanism achieves the near-optimal tradeoff between bid ratio condition and fair value within a restricted class of mechanisms. In Section \ref{subsec:comparison}, we show that the fair value achieved by this mechanism is close to the upper bound established in Section \ref{subsec:uniform} for allocation mechanisms with a uniform metric. We emphasize that while our negative result in Section \ref{subsec:uniform} applies to general online algorithms which satisfy multiple-task fairness with access to the underlying metric, our positive result applies to a mechanism that is metric-oblivious and satisfies the stronger notion of total variation fairness. All of the proofs can be found in the Appendix.


\subsection{Fairness in bids}
\label{subsec:bidratio}


Our goal is to develop an allocation mechanism that simultaneously satisfies total variation fairness and achieves large fair value with respect to the Unfair-OPT. As one might expect, it is impossible to achieve this if advertisers are allowed to place arbitrary bids on users without regard to the relevant similarity metric over users.\footnote{One could consider platform mechanisms which alter or decrease the bids of advertisers as needed to achieve fairness constraints but this has the downside of (1) making the platform responsible for advertiser behavior and (2) making it difficult for advertisers to optimize their bids.} The question thus becomes: what kind of fairness constraint on bids enables a reasonable fair value? The following example illustrates the need for a fairness constraint that requires an advertiser's bids on pairs of similar users to be close in ratio, even when allocations are merely required to satisfy the weaker condition of multiple-task fairness. In particular, being close in terms of their absolute difference is not sufficient to achieve a good fair value.
\begin{example}
\label{example:perfectmatching}
Suppose that there are $k$ advertisers and exactly $k$ users. Suppose that the metric is uniform: for some parameter $d \in [0,1]$, every pair of users is a distance of $d$ apart. For each $i\in [k]$, advertiser $i$ bids $b^{\text{high}}$ on user $i$ and $b^{\text{low}} (< b^{\text{high}})$ on all other users $j\ne i$. Observe that Unfair-OPT $= k b^{\text{high}}$. On the other hand, due the symmetry of this instance and the fact that a fair allocation requires that $|p_i^i -p_i^j|\le d$, for each user, it turns out the optimal fair allocation assigns an allocation probability of $(1-d)/k$ to all advertisers with the low bid and a probability of $(1-d)/k + d$ to the advertiser with the high bid. The fair value of this allocation turns out to be\footnote{We prove this explicitly in Lemma \ref{lemma:exampleofflineapproximationratio} in Appendix \ref{subsubsec:offline} as a corollary of a more general result about the optimal utility achieved by a fair offline mechanism on a uniform metric.}
$d+ \frac{1-d}{k} + (1-d)\frac{(k-1)}{k} \frac{b^{\text{low}}}{b^{\text{high}}}$.
Observe that a fair value of $d+(1-d)/k$ is trivial to achieve via a fair allocation.\footnote{In particular, for each user, assigning an allocation probability of $(1-d)/k + d$ to the highest bidder and $(1-d)/k$ to all other advertisers achieves this bound.} When $d$ is very small, this trivial bound is tiny. If $b^{\text{high}}\gg b^{\text{low}}$, then, in this example, no fair allocation can perform much better than the trivial algorithm. In order to be able to do better, $b^{\text{high}}/b^{\text{low}}$ must be bounded.
\end{example}

Motivated by the example above, we require that for every advertiser and every pair of users, the ratio of the bids the advertiser places on the users is bounded by a function of the distance between the users according to $\bfd$. The closer the two users, the closer this ratio bound should be to $1$; on the other hand, the ratio bound should be large between far apart users. We formally define this constraint as follows.
\begin{definition}
A {\bf bid ratio constraint} is a function $f:[0,1] \rightarrow [1, \infty]$. We say that the bid function $b^i$ of advertiser $i$ satisfies the bid ratio constraint $f$ with respect to metric $\bfd$ if we have for all $u,v\in U$: $\frac{1}{f(\bfd(u,v))} \le \frac{b^i_u}{b^i_v} \le f(\bfd(u,v))$.
\end{definition}

How should we choose $f$?
On the one hand, the bid ratio constraint needs to be sufficiently strict to provide meaningful fairness and utility guarantees and on the other hand it cannot be so overly restrictive as to prohibit reasonably expressive bidding strategies.
We characterize $f$ by boundary conditions for identical users and maximally distant users with the requirement that $f$ is weakly increasing and $f>1$ in the intermediate range. In the case of identical users, i.e. $\bfd(u,v)=0$, the advertiser is required to place identical bids, i.e., $f(0)=1$. For maximally distant users, i.e. $\bfd(u,v)=1$, we choose $f(1) = \infty$, allowing advertisers to bid arbitrarily differently on this pair. For ease of analysis, we also make $f$continuous.

\begin{figure}
    \centering
    \includegraphics[scale=0.6]{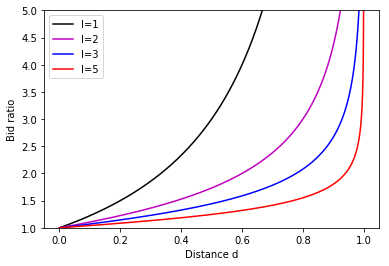}
    \caption{Bid ratio condition $f_l(d) = \left(\frac{1+d}{1-d}\right)^{1/l}$}
    \label{fig:bidratioshape}
\end{figure}

In this work, we show that a specific parameterized class of bid ratio constraints that satisfy the above properties performs quite well in terms of fair value. The family is parameterized by $l\ge 1$, and is defined as:
$f_l(d) = \left(\frac{1+d}{1-d}\right)^l$.
Figure~\ref{fig:bidratioshape} displays some functions in this family. Note that as the parameter $l$ increases, the bid ratio condition becomes more and more strict. In Appendix \ref{sec:convexitybidratio}, we discuss implications of certain structural properties of $f_l(d)$.

We emphasize that bid ratios are not required to satisfy the constraint exactly, but rather should lie \textit{at or below} the imposed curve. From an algorithmic viewpoint, this means that if we design an algorithm based on the polynomial family described above but the actual bid ratio constraint imposed on bids, say $g$, does not belong to this family, it nevertheless suffices for the algorithm to find a value of the parameter $l$ for which $f_l(d)\ge g(d)$ for all $d\in [0,1]$ and use the function $f_l$ in making allocations. 


\subsection{Lower bounds on fair value using uniform metrics}\label{subsec:uniform}
We prove upper bounds on the fair value of any allocation mechanism that satisfies multiple-task fairness. Observe that these upper bounds apply also to total variation fairness, which is a stronger requirement. We use uniform metrics, i.e. metrics of the form $\bfd(u,v) = d$ for all $u\ne v$. 
We use Example \ref{example:perfectmatching} to show an upper bound on the fair value as a function of the bid ratio constraint.\footnote{In Appendix \ref{sec:uniformmechanisms}, we give an explicit formula for the optimal revenue that can be achieved in the offline setting for a uniform metrics as a function of the bids and $d$. This formula yields the bound in Lemma \ref{lemma:offlineexamplerevenue} on Example \ref{example:perfectmatching}.} 
\begin{lemma}
\label{lemma:offlineexamplerevenue}
Given $d\in [0,1]$ and $\alpha=f(d)$, there exists an instance of the online advertising problem for which the fair value of every offline allocation mechanism satisfying multiple-task fairness with respect to the uniform metric $\bfd(u,v) = d$ is at most 
$d+\frac{1-d}{k} + \frac{(1-d)(k-1)}{k \alpha}  \le \frac{1}{k} + \frac{1}{\alpha} + d$. 
\end{lemma} 
\noindent Observe that as $d$ increases, the upper bound on the fair value increases, due to a weaker fairness constraint on setting allocation probabilities. On the other hand, for any fixed $d$ and $k$, the upper bound decreases as a function of $\alpha$: as $\alpha$ increases, weakening the fairness constraint on bids, the algorithm's performance becomes worse.


In the online setting it is possible to prove even stronger bounds on the fair value.\footnote{To be explicit, the online lower bound applies in the limit as $|U| \rightarrow \infty$, where the adversary is given access to the probabilities output by the mechanism when designing the next set of bids.} More specifically, we can tighten the $1/\alpha$ term in the fair value to $1 / \alpha^2$. 
In Appendix \ref{sec:uniformmechanisms}, we develop online allocation mechanisms that are tailored to uniform metrics and achieve a fair value that nearly matches these lower bounds, demonstrating that stronger lower bounds for general online algorithms cannot be obtained through uniform metrics.
\begin{lemma}
\label{lemma:genlowerboundbidratio}
Given $d\in [0,1-1/k]$ and $\alpha=f(d)$, there exists an instance of the online advertising problem for which no online allocation mechanism satisfying multiple-task fairness with respect to the uniform metric $\bfd(u,v) = d$ can obtain fair value better than $\alpha^{-2}\left(1 - \frac{1}{k} -d \right) + \frac{1}{k} + d \le \frac{1}{k} + \frac{1}{\alpha^2} + d$. 
\end{lemma}
The main idea of the proof of Lemma \ref{lemma:genlowerboundbidratio} is to make the bids on the first user equal. Then for the next user, the adversary maximally increases some bids and decreases other bids so that the advertiser receiving the lowest probability on the first user has the highest bid on the second user. The distance metric limits the extent to which the mechanism can increase the probability placed on this advertiser for the second user due to the low probability placed on the first user.

\subsection{Mechanisms for the general metric case}\label{subsec:gensimilar}
We construct an allocation mechanism for a general fairness metric $\bfd$ that achieves total variation fairness and a large fair value. From a mechanism design standpoint, it is desirable for our mechanism to satisfy certain other properties. Most important is \textit{metric-obliviousness}, i.e. where the mechanism doesn't require direct knowledge of $\bfd$. In addition, the following other properties are desirable:
\begin{enumerate}[leftmargin=*]
    \item \textit{Identity-obliviousness}: that is, the mechanism treats advertisers in a symmetric manner in each individual iteration --- the allocation to an advertiser does not depend on their identity. 
    \item \textit{Monotonicity}: that is, the allocation probabilities increase monotonically as functions of the advertisers' bids, which means that we can make this mechanism truthful by setting the payoffs appropriately by Myerson's lemma.
    \item \textit{History-obliviousness}, which has the nice implication that the memory required by the mechanism is independent of the number of users and the solution is \textit{independent} of the ordering of the users so we don't need to worry about impact of the order in which the users are given on advertiser strategies and/or utility. 
    \item \textit{Protecting against advertiser splitting}: We would like the mechanism to disincentivize an advertiser from splitting up into sub-advertisers (or submitting multiple ads) in an attempt to obtain a higher probability allocation.
\end{enumerate}

We now describe a class of mechanisms that satisfy these properties. The key intuition is to convert the bids on a user into probabilities using a function that places higher probabilities on higher bids. Note that the first three properties (identity-oblivious, monotonic, and history-oblivious) mean that the mechanism must be defined by a symmetric, coordinate-wise increasing function $G: \left(\mathbb{R}^{\ge 0}\right)^k \rightarrow \left\{(p^1, \ldots, p^k) \mid p^i \ge 0, \sum_{i=1}^k p^i \le 1 \right\}$ that maps bids into probabilities. We observe that the overall optimal solution (Unfair-OPT) can be placed in this framework: the mechanism that distributes the probability mass equally among the highest bidders for each user corresponds to the function that places the full mass on the highest bids. This function can be viewed as assigning allocation probabilities in proportion to their contribution to the 
$\ell_{\infty}$-norm over the bids. 

\subsubsection*{Proportional allocation mechanisms}
One rich class of mechanisms of this form are mechanisms where the probabilities are proportional to some deterministic function $g$ of the bids, i.e. where $p^i_u \propto g(b^i_u)$, with appropriate normalization to make $\sum_{i=1}^k p^i_u = 1$. We call these mechanisms proportional allocation mechanisms, defined as follows:\footnote{The special case of the proportional allocation mechanism with $g(x) = x$ was considered in a different context in \cite{AZM2018}.} 
\begin{mech}
\label{mech:propalloc}
Let $g: \mathbb{R}^{\ge 0} \rightarrow \mathbb{R}^{\ge 0}$ be a continuous, super-additive (i.e. $g(x) + g(y) \le g(x+y)$), increasing function. The \textbf{proportional allocation mechanism with parameter $g$} assigns $p^i_u = \frac{g(b^i_u)}{\sum_{j=1}^k g(b^i_u)}$ for every user $u\in U$ and advertiser $i\in [k]$.
\end{mech}
\noindent It is straightforward to verify that proportional allocation mechanisms are identity-oblivious, monotonic, history-oblivious, and protect against submitting multiple ads (this last property follows from the super-additivity of $g$). These mechanisms bear similarity to position auctions\footnote{Roughly speaking, a position auction \cite{Varian2006, EOS2005} only uses the ordering of the bids (and not the bid values or advertiser identities) in determining the allocation. See Appendix \ref{sec:positionauctions} for a discussion of why position auctions cannot achieve both fairness and a high fair value.}: while proportional allocation mechanisms are not position auctions since they can still rely heavily on the bids, they are ``close'' to position auctions in that the highest bidder is always assigned the highest probability regardless of their identity.

\subsubsection*{A proportional allocation mechanism with high fair value}
We construct a family of functions $g$ where the fair value of the proportional allocation mechanisms is high. More specifically, we show that $g(x) = x^l$ for $l \ge 1$ can achieve fair value approaching $1$ as $l \rightarrow \infty$. This mechanism can be viewed as assigning allocations in proportion to each bid's contribution to the $\ell_l$-norm of the bid vector. We emphasize that our bound on the fair value does not require bids to satisfy the bid ratio constraint.
\begin{theorem}
\label{thm:revenuebound}
Let $M$ be a proportional allocation mechanism with parameter $g(x) = x^l$ for a positive integer $l$. If $k \ge 9$ and $l \ge 1$, then the fair value of $M$ is at least $(k-1)^{-1/l}\frac{k-1}{k} + \frac{1}{k}$. 
\end{theorem}
\noindent Observe that fair value is an increasing function of $l$, and as $l \rightarrow \infty$ (where $M$ places the entire mass on the highest bid), the bound equals $1$ as expected. This means that fair value can be made arbitrarily close to $1$ by sufficiently strengthening the bid ratio constraint. To achieve a fair value of $r$, we can set $l \ge \frac{\log(k-1)}{\log(1/r)}$. 

We show that with the bid ratio condition $f_l(d) = \left(\frac{1+d}{1-d} \right)^{1/l}$ shown in Figure \ref{fig:bidratioshape}, this mechanism satisfies total variation fairness.
\begin{theorem}
\label{thm:bidratio}
Let $M$ be a proportional allocation mechanism with parameter $g(x) = x^l$ for a positive integer $l$. If all advertisers in a category satisfy the bid ratio condition $f_l(d) = \left(\frac{1+d}{1-d} \right)^{1/l}$, $M$ satisfies total variation fairness in that category.  
\end{theorem}
\noindent Observe that the bid ratio condition becomes stronger as $l$ increases. Thus, for a fixed number of advertisers, a higher fair value is accompanied by a stronger bid ratio condition. Moreover, observe that to maintain a fair value of $r$, a stronger bid ratio condition is needed as the number of advertisers increases, since $l = \frac{\log(k-1)}{\log(1/r)}$ grows with $k$. In Appendix \ref{sec:identicalrelaxations} we discuss relaxations of the fairness constraint and analyze the bid ratio condition needed for Mechanism \ref{mech:propalloc} with parameter $x^l$ under those relaxations.

\subsubsection*{Near-optimality within proportional allocation mechanisms}
A natural question is to ask is whether Mechanism \ref{mech:propalloc} with a different function $g$ can achieve a a much better fair value, potentially using a differently shaped bid ratio condition. In fact, we show that Mechanism \ref{mech:propalloc} is nearly optimal within the family of proportional allocation mechanisms. We specifically prove that \textit{any} proportional allocation mechanism achieving a certain fair value will have a corresponding bid ratio condition that is \textit{point-wise} stronger than $f_l(d) = \left(\frac{1+d}{1-d} \right)^{1/l}$, where $l$ within a constant factor of what is achieved by Mechanism \ref{mech:propalloc} with $g(x) = x^l$. This result demonstrates that changing the \textit{shape} of the bid ratio condition  will not significantly improve the fair value. We show the following lower bound: 
\begin{lemma}
\label{lemma:mainpropalloclowerbound}
Suppose that $M$ is a proportional allocation mechanism achieves fair value $r$ and achieves total variation fairness with a bid ratio condition of $f$. For $l = \frac{\log(k-1)}{2 \log(1/r)} - 0.5$, we have that 
$f(d) \le f_l(d) = \left(\frac{1+d}{1-d} \right)^{1/l}$  for infinitely many points on $f$ (more specifically, for $d \in D := \left\{ d \mid f(d) = \left(\frac{1}{r^2} \right)^{m}, m \in \left\{n, \frac{1}{n} \mid n \in \mathbb{N}\right\}\right\}$).
  \end{lemma}
\noindent How does the lower bound in Lemma \ref{lemma:mainpropalloclowerbound} compare to the proportional allocation mechanisms with parameter $x^l$? Lemma \ref{lemma:mainpropalloclowerbound} shows that any proportional allocation mechanism will satisfy $f(d) \le f_l(d)$ where $l = \frac{\log(k-1)}{2 \log(1/r)} - 0.5$ for infinitely many points on the curve. Meanwhile, Theorem \ref{thm:revenuebound} and Theorem \ref{thm:bidratio} show that to achieve a fair value $r$, it suffices to take $f_l(d)$ with $l = \frac{\log(k-1)}{\log(1/r)}$. Thus, there is essentially a constant factor difference in the lower and upper bounds on $l$. 

A consequence of Lemma \ref{lemma:mainpropalloclowerbound} is in the family of proportional allocation mechanisms, the fair value must necessarily degrade with the number of advertisers $k$ if the bid ratio condition is fixed. Or equivalently, to maintain a fair value of $r$, a stronger bid ratio condition is needed as the number of advertisers increases. In fact, Lemma \ref{lemma:mainpropalloclowerbound} implies that that the proportional mechanism with parameter $x^l$ achieves the optimal rate of change of the bid ratio condition as a function of $k$: within the family of proportional allocation mechanisms, a better dependence on $k$ is not possible.

\subsection{Discussion}
\label{subsec:comparison}
\begin{figure}
    \centering
    \begin{subfigure}[b]{0.225\textwidth}
    \centering
    \includegraphics[scale=0.3]{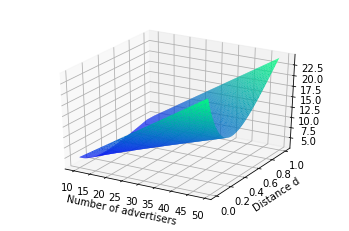}
    \caption{$l=1$}
    \end{subfigure}
    \begin{subfigure}[b]{0.225\textwidth}
    \centering
    \includegraphics[scale=0.3]{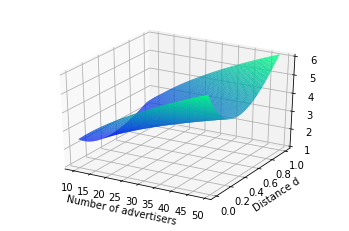}
    \caption{$l=2$}
    \end{subfigure}
    \quad
    \begin{subfigure}[b]{0.225\textwidth}
    \centering
    \includegraphics[scale=0.3]{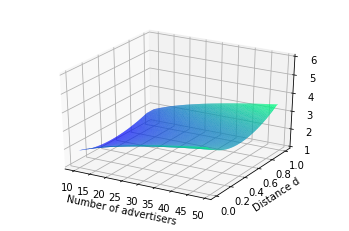}
    \caption{$l=3$}
    \end{subfigure}
    \begin{subfigure}[b]{0.225\textwidth}
    \centering
    \includegraphics[scale=0.3]{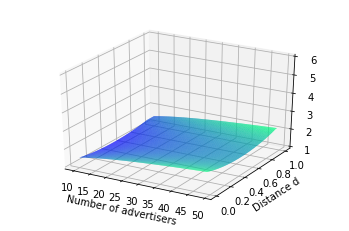}
    \caption{$l=5$}
    \end{subfigure}
    \setlength{\belowcaptionskip}{-10pt}
    \caption{Illustrations of the ratio of upper and lower bounds on fair values for different values of $l$. Note that the $z-$axis scale for $l=1$ differs from the others.}
    \label{fig:plots}
\end{figure}

In Section \ref{subsec:gensimilar}, we showed that Mechanism \ref{mech:propalloc} with $g(x) = x^l$ is nearly optimal in the class of proportional allocation mechanisms. Now, we compare Mechanism \ref{mech:propalloc} with $g(x) = x^l$ to general online mechanisms, using our negative results in Section \ref{subsec:uniform}. It is a little tricky to directly compare the fair value lower bounds achieved by Mechanism \ref{mech:propalloc} with $g(x) = x^l$ with the upper bounds in Section \ref{subsec:uniform}, because the bounds depend on different parameters. In particular, the lower bounds hold for arbitrary metrics whereas the upper bounds are designed only for the uniform metric. To perform an apples-to-apples comparison, we fix parameters $k$, $l$, and some number $d\in (0,1)$, and set $\alpha=f_l(d)$. Figure \ref{fig:plots} displays the ratio of the upper bound and lower bound for various parameter settings. Observe that the ratio is bounded by a reasonably small constant except when $l$ is very small. This indicates that Mechanism \ref{mech:propalloc} with parameter $g(x) = x^l$ in general obtains a large fraction of the utility that can be obtained by any online mechanism satisfying multiple-task fairness, despite satisfying a stronger form of fairness and nice mechanism design properties.

\section{Inter-category fairness}\label{sec:different}
\newcommand{\bfd}{{\bf d}}
In this section, we consider the setting where different product categories correspond to very different metrics. We specifically consider the setting where every category has exactly one advertiser (i.e. where $c = k$). We first show it is not possible to achieve the same tradeoff between multiple-task fairness and utility as in the case of identical metrics. We then show that with inter-category envy-freeness, significantly better tradeoffs are possible. 

In Section \ref{subsec:firstprice}, we focus on the Unfair-OPT benchmark considered in the previous section. First, we show that multiple-task fairness is suboptimal for users, advertisers, and the platform. We then consider the weaker notion of inter-category envy-freeness introduced in Section \ref{sec:definitions}, and we present tight upper and lower bounds on the fair value achievable as a function of upper bounds on the sizes of the preferred sets. In Section \ref{subsec:relaxedrevenue}, we argue in favor of a weaker benchmark to evaluate the performance of fair allocation mechanisms, and show that this benchmark can be exactly met by mechanisms that satisfy inter-category envy-freeness. Proofs are in the Appendix.


\subsection{Inter-category envy-freeness and fair value relative to Unfair-OPT}\label{subsec:firstprice}
First, we show a upper bound that demonstrates that multiple-task fairness is in conflict with utility, even in the offline setting, if the metrics are permitted to be different. The result is based on the ``jack of all trades'' example in the introduction. 
Formally:
\begin{example}[Jack-of-all-trades]
\label{example:wantedperson}
Suppose that the universe has $c+1$ users $u_1, \ldots, u_{c+1}$ and there are $c$ categories (with one advertiser per category). The metric $\bfd^i$ is defined so that $\bfd^i(u_{c+1}, u_i) = 0$ and all other distances are $1$. Suppose that advertiser $i$ bids $1$ on $u_i$ and $u_{c+1}$, and $0$ on everybody else. Observe that the bids are fair.\footnote{We assume that $f(1) = \infty$.} 
\end{example}

Consider any allocation mechanism and suppose that this mechanism chooses an ad from distribution $(q_1, \cdots, q_c)$ to display for user $u_{c+1}$. Then, in order to respect multiple task fairness as defined above, the mechanism cannot allocate ad $i$ to user $u_i$ with probability greater than $q_i$. As a result, most of the ``specialists'' necessarily obtain a low allocation within their desired categories: since the ``jack of all trades'' can only be served a single ad, the ``specialists'' are penalized. 
Multiple-task fairness thus limits the allocation of almost all of the users.  Moreover, the utility of any fair allocation is bounded by some constant, whereas an unfair allocation can achieve utility $c+1$. We obtain the following bound on fair value: 

\begin{proposition}
\label{prop:lowerboundofflinedifferent}
Suppose that bids satisfy the bid ratio constraint $f$, then no offline mechanism that satisfies multiple-task fairness across $c$ categories can obtain a fair value more than $\frac{2}{c+1}$.
\end{proposition}

We now consider the weaker fairness notion of inter-category envy-freeness defined in Section~\ref{sec:definitions} that limits the number of fairness constraints imposed by any given user. We assume that the platform obtains from each user $u\in U$ a preferred set $S_u\subseteq [c]$ of categories as the user arrives (i.e. the preferences are not known to the allocation mechanism in advance). Inter-category envy-freeness requires that the total allocation of ads in $S_u$ to the user $u$ should be at least as large as the total allocation of ads in $S_u$ to any other user. Observe that our definition of fairness doesn't involve a metric over the users. 
Moreover, obtaining good utility requires that we place a reasonable amount of mass on the highest bid(s) for the user $u$. 

We develop an allocation mechanism that reserves some allocation mass for the category with the highest bid on each user $u$, and distributes the remaining allocation probability across categories in $S_u$. In doing so, we must ensure that no subset of categories gets too much mass in total. This is because if such a set $S$ exists, and a future user $v$ sets $S_v = S$, then the mechanism is forced to give a high allocation to this set for user $v$. Unless $S$ is small, this may then leave little probability mass for the highest bidder for $v$.


We now formally define our mechanism and bound the fair value achieved by it. 
\begin{mech}
\label{mech:userspecified}
The {\bf equal-spread mechanism with parameters $\beta$ and $C$} is defined as follows. We assume that every user $u$ specifies a subset $S_u \subset [c]$ with $|S_u| \le C$. Let $C_u = \operatorname{argmax}_{1 \le j \le c}\{b_u^j\}$ be a category with the highest bid. The mechanism assigns an allocation probability of $p_{\text{high}}$ to $C_u$ and $p_{\text{fairness}}$ to categories in $S_u \setminus \left\{C_u\right\}$, where $p_{\text{high}} = \frac{1}{1 + \beta^{-1} + \ldots + \beta^{-C}}$ and $p_{\text{fairness}} =  \frac{\beta^{-1} + \ldots + \beta^{-|S_u|}}{(|S_u|)(1 + \beta^{-1} + \ldots + \beta^{-C})}$. 
\end{mech}



The parameter $\beta\ge 1$ allows the mechanism to trade-off between fairness and utility. By allowing the mechanism to achieve $\beta$-inter-category envy-freeness for larger $\beta$ values, we obtain a better approximation to the Unfair-OPT. 

\begin{theorem}
\label{thm:upperbounduserspecified}
If every user's preferred set contains $\le C$ categories, then for any $\beta\ge 1$, the equal-spread mechanism with parameters $\beta$ and $C$ (Mechanism \ref{mech:userspecified}) satisfies $\beta$-inter-category envy-freeness and achieves a fair value of $\ge \frac{1}{1 + \beta^{-1} + \ldots + \beta^{-C}}$.
\end{theorem}

We show a matching upper bound on the fair value, thus showing that Mechanism \ref{mech:userspecified} is optimal. Our proof boils down to bounding the amount of mass that can be placed on the highest bid. 
We construct a sequence of users with the property that bids in $S_u$ are always $0$ and each user has a category $C_u \not\in S_u$ that bids $1$. We adaptively construct the sets $S_u$ and $C_u$ to minimize the fair value.

\begin{lemma}
\label{lemma:lowerbounduserspecified}
Suppose that $C< c$, and every user's preferred set can contain any number of categories $\le C$. Then, regardless of the bid ratio constraint $f$ imposed on the advertisers (but assuming $f(1)=\infty$), any online mechanism that satisfies $\beta$-inter-category envy-freeness obtains a fair value of at most $\frac{1}{1 + \beta^{-1} + \ldots + \beta^{-C}}$.
\end{lemma}

In the above construction, we consider a uniform metric with distance $1$ (where the bid ratio conditions do not implicitly provide any guarantees). Nonetheless, even though the users are maximally distant, the inter-category envy-freeness still provides strong uni-dimensional guarantees between these users. A natural question to ask is: can we achieve a higher fair value by considering a relaxed version of inter-category envy-freeness that reintroduces the metric? However, we show in Appendix \ref{sec:differentmetricslowerbounds} that even with these relaxations, it is not possible to achieve a much higher fair value than that achieved by Mechanism \ref{mech:userspecified}. 



\begin{figure}
    \centering
    \includegraphics[scale=0.7]{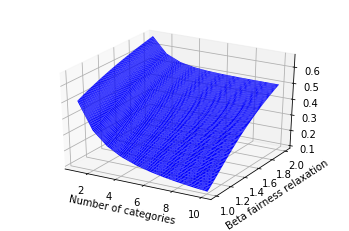}
    \setlength{\belowcaptionskip}{-10pt}
    \setlength{\abovecaptionskip}{-5pt}
    \caption{Fair value for inter-category envy-freeness}
    \label{fig:3dplotuserdirected}
\end{figure}

We plot the tight bound on fair value obtained for inter-category envy-freeness as a function of $\beta$ and $C$ in Figure \ref{fig:3dplotuserdirected}. For the dependence on $\beta$, observe that as $\beta$ increases, the weakened fairness guarantees cause the fair value to increase. For the dependence on $C$, observe that the mechanism must balance between allocating to a category with the highest bid to achieve high utility, and allocating to categories in $S_u$ to achieve inter-category envy-freeness. As $C$ increases, the mechanism has a greater number of categories to consider for each user (and the highest bid can still be outside of $S_u$), thus causing the optimal fair value to decrease. Let's now consider the strongest setting of $\beta = 1$, so the fair value becomes $\frac{1}{C+1}$. The fair value thus becomes small when users are permitted to specify a large number of categories, and when $C = 1$ (and $\beta = 1$), the fair value is $1/2$. 

\subsection{Relaxing the utility benchmark}\label{subsec:relaxedrevenue}
The results in the previous section, for the strongest form of fairness, place an upper bound of $1/2$ (or lower) on the fair value that can achieved by inter-category envy-freeness. In this section, we consider the setting where we restrict to mechanisms that receive utility only for allocations in $S_u$: that is, the utility is $\sum_{u \in U} \sum_{i \in S_u} p_u^i b_u^i$. In this case, we assume that user-specified categories are aligned with interest, and in a click-through-rate based revenue/utility model, the platform only obtains benefit from allocations within the user-specified sets. 

It is straightforward to see that the best possible utility achieved by any (potentially unfair) mechanism in this restricted class is $\sum_{u \in U} \max_{i \in S_u} b^i_u$, which may be much smaller than unrestricted optimal utility of $\sum_{u \in U} \max_{i \in [c]} b^i_u$. Thus, when considering the utility for this setting, we compare against $\sum_{u \in U} \max_{i \in S_u} b^i_u$, the best possible user-preference-compatible utility for a mechanism in this setting. This motivates considering the relaxed fair value given by: $\frac{\sum_{u \in U} \sum_{i \in S_u} p^i_u b^i_u}{\sum_{u \in U} \max_{i \in S_u} b^i_u}$. Note that when there can be more one advertiser per category, this quantity naturally generalizes to $\frac{\sum_{u \in U} \sum_{i \in S_u} \sum_{j \in C_i} p^j_u b^j_u}{\sum_{u \in U} \max_{i \in S_u, j \in C_i} b^j_u}$.

Returning to the case where there is at most one advertiser per category, we show a strong positive result in this setting. More specifically, we remark that a simple highest-bidder-wins mechanism achieves inter-category envy-freeness and a relaxed fair value of $1$. This mechanisms does not assume that all users specify the same number of categories and permits users to specify any number of categories. 
\begin{mech}
\label{mech:firstpricevariant}
For each $u\in U$, the mechanism allocates a probability of $1$ to the category in $S_u$ with the highest bid. If there are multiple categories tied for the highest bid, then the mechanism splits the probability equally between these categories. 
\end{mech}
\begin{theorem}
\label{thm:userdirectedpositiveresult}
Mechanism \ref{mech:firstpricevariant} achieves inter-category envy-freeness with $\beta = 1$ and achieves a relaxed fair value of $1$.
\end{theorem}
\noindent 
Given the simplicity of Mechanism \ref{mech:firstpricevariant}, a natural question is to ask is: can we obtain better fairness guarantees while still maintaining a high relaxed fair value? However, in Appendix \ref{sec:multipletaskrelaxed}, we show that it is not possible to have multiple-task fairness guarantees and a high fair value in this setting even against the relaxed benchmark.

\section{Compositional fairness}\label{sec:combo}
\newcommand{\bfp}{{\bf p}}

\newcommand{\bfd}{{\bf d}}
 
We now discuss how to combine our mechanisms to handle the setting where there can be multiple categories and multiple advertisers in each category. Let $\{C_1, \ldots, C_c\}$ denote a partition of the set $[k]$ of advertisers into $c$ categories. We can compose our mechanisms from Section \ref{sec:identical} and Section \ref{sec:different} by running mechanisms from Section \ref{sec:different} to allocate probability between categories, and mechanisms from Section \ref{sec:identical} to distribute the probability assigned to a category between advertisers in that category. 
The key intuition for fairness is that compositional fairness is implied by inter-category envy-freeness and intra-category total variation fairness.\footnote{In order to conclude that $\sum_{i \in C_j} p^i_u$ is actually the probability that the first mechanism assigned to category $C_l$, we require that the second mechanism assigns the full $1$ probability mass to each user. This is true of proportional allocation mechanisms (Mechanism \ref{mech:propalloc}).} 
Moreover, we show that these combined constructions achieve a high fair value, since the fair value of the composed mechanism is the product of the fair values of the individual mechanisms. Note that the two choices of category selection mechanisms in Section \ref{sec:different} result in combined constructions that differ in terms of both utility guarantees (i.e. choice of benchmark) and category distribution properties (i.e., concentrating or spreading probability between categories), which are important for practical considerations. Proofs are in the Appendix. 

\subsection*{High fair value compared to the relaxed benchmark}
To achieve a high fair value compared to the relaxed utility benchmark (taking user preferences as an indicator of click probability discussed in Section \ref{subsec:relaxedrevenue}), we can compose Mechanism \ref{mech:firstpricevariant} with Mechanism \ref{mech:propalloc}. The idea is that we run Mechanism \ref{mech:firstpricevariant} to identify the category in $S_u$ with the highest bid, and then we run Mechanism \ref{mech:propalloc} to divide the probability mass between advertisers in this category.
\begin{mech}
\label{mech:composed1}
For each user $u$, the mechanism runs Mechanism \ref{mech:firstpricevariant} to allocate probabilities between categories, where for $1 \le j \le c$, the bid $B_u^j$ is taken to be $\max_{i \in C_j} b_u^i$.\footnote{This sets the bid for a category to be the highest bid of any advertiser within that category.} The mechanism then determines the conditional probabilities for advertisers within each category using Mechanism \ref{mech:propalloc} with parameter $g(x) = x^l$. 
\end{mech}
\begin{theorem}
\label{thm:composition1}
Let $k' = \max_{1 \le j \le c} |C_j|$ be the maximum number of advertisers in any category. If all advertisers $1 \le i \le k$ satisfy the bid ratio condition $f^i(d) \le \left(\frac{1+d}{1-d} \right)^{1/l}$, then Mechanism \ref{mech:composed1} achieves compositional fairness and a relaxed fair value of at least $(k'-1)^{-1/l} \frac{k'-1}{k'} + \frac{1}{k'}$. 
\end{theorem}

\subsection*{High fair value compared to Unfair-OPT}
It is likewise possible to combine mechanisms \ref{mech:propalloc} and \ref{mech:userspecified} to obtain bounds on the fair value against Unfair-OPT.
\begin{mech}
\label{mech:composed2}
For each user $u$, the mechanism runs Mechanism \ref{mech:userspecified} with parameters $\beta$ and $C$ to allocate probabilities between categories, where for $1 \le j \le c$, the bid $B_u^j$ is taken to be $\max_{i \in C_j} b_u^i$.\footnote{Again, this sets the bid for a category to be the highest bid of any advertiser within that category.} The mechanism then determines the conditional probabilities for advertisers within each category using Mechanism \ref{mech:propalloc} with parameter $g(x) = x^l$.
\end{mech}
\begin{theorem}
\label{thm:composition2}
Suppose that every user's preferred set contains at most $C$ categories. Let $k' = \max_{1 \le j \le c} |C_j|$ be the maximum number of advertisers in any category. If all advertisers $1 \le i \le k$ satisfy the bid ratio condition $f^i(d) \le \left(\frac{1+d}{1-d} \right)^{1/l}$, then Mechanism \ref{mech:composed2} achieves compositional fairness and a fair value of at least $\left(\frac{1}{1 + \beta^{-1} + \ldots + \beta^{-C}} \right)\left((k'-1)^{-1/l} \frac{k'-1}{k'} + \frac{1}{k'}\right)$. 
\end{theorem}

\section{Future work}\label{sec:future}
In this work, we give an initial framework for understanding the utility and fairness properties of a combination of individual fairness and envy-freeness. We highlight areas for future work below. 
\begin{itemize}[leftmargin=*]
\item \textit{Incentivizing and auditing fair bidding.} A significant benefit of the online mechanism presented for intra-category competition is that it is \textit{metric oblivious} which frees the platform from explicitly checking whether bids placed are in accordance with the relevant fairness metric. Although total variation fairness does mitigate some malicious advertiser behavior, there is still an open question of how best to incentivize, audit, and enforce fair bidding. 
We anticipate a mechanism which can impose penalties (either monetary penalties or reduced platform access) on advertisers who do not behave fairly may provide sufficient incentive to follow the rules.

\item \textit{Closing the gaps and tighter lower bounds.} 
Although uniform metrics provide a good starting point for lower bounds, and we have provided nearly matching bounds for proportional allocation mechanisms, we anticipate that it may be possible to show more general lower bounds in certain oblivious settings, i.e., requiring metric or identity obliviousness.


\item \textit{Multiple slot auctions.} In this work, we have focused on the case of single slot auctions. However, in practice multiple slot auctions, or auctions where each user appears at several times, are more common.
In addition to technical implementation questions, there are several important considerations in defining fairness for the setting. For instance, the ordering of advertisements may be important when there is significant visual distinction (e.g. the user must scroll to see ads lower in the slate).
\item \textit{Different combinations of inter- and intra-category fairness.} We have presented one combination of definitions of inter- and intra-category fairness, but it's likely that others (e.g., intra-category total variation fairness and inter-category PIIF) may have interesting properties.  
\end{itemize}

\bibliographystyle{plain}
\bibliography{bibliography.bib}

 
 
\appendix

\section{Proofs for Section 4}

\subsection{Proofs for Section \ref{subsec:uniform}}
In Section \ref{subsubsec:offline}, we prove bounds for offline mechanisms. In Section \ref{subsubsec:online}, we prove bounds for online mechanisms. 
\subsubsection{Offline mechanisms}\label{subsubsec:offline}
First, we compute the optimal offline mechanism revenue for uniform metrics.
\begin{lemma}
\label{result:offlineregularrevenue}
Suppose that $A_x$ is an advertiser such that 
\[\sum_{u \in U} b^x_u = \max_{1 \le i \le k} \left(\sum_{u \in U}b^i_u\right).\] 
If the metric is uniform with distance $d$, then the optimal offline, multiple-task fair mechanism achieves a revenue of exactly:
\[\max(0, 1-md)\sum_{u \in U} b_u^x + d\sum_{i=1}^{m-1} \text{ith price auction revenue}\]
\[+ (\min(d, 1 - (m-1)d)) \text{mth price auction revenue}.\]
where $m = 0$ if $\sum_{u \in U}  b^x_u$ is the first price auction revenue, and otherwise, 
$m$ is the maximum integer in $[1, \min(k-1,\frac{1}{d} + 1)]$ such that the mth price auction revenue is bigger than $\sum_{u \in U}  b^x_u$. 
\end{lemma}

The proof for this lemma boils down to solving the following linear program. In the offline setting for multiple-task fairness, an optimal-revenue mechanism that achieves multiple-task fairness is solving the following linear program, where the solutions are $\left\{p_u^i \right\}_{1 \le i \le k, u \in U}$, and the rest of the variables are inputs. 
\begin{align*}
    &\max \left(\sum_{i=1}^k \sum_{u \in U} b^i_u p^i_u \right) \text{ such that:} \\
    &\sum_{i=1}^k p^i_u \le 1 \text{ for all } u \in U \\
    &p^i_u - p^i_v \le d^i(u,v) \text{ for all } u \neq v \in U \\
    &p^i_u \ge 0 
\end{align*} 

We first show the following property of the LP.
In order to better study this LP, we also consider the dual with variables $\left\{z_u\right\}_{i \in U}$ and $\left\{w^i_{u,v}\right\}_{1 \le i \le k, u \neq v \in U}$:
\begin{align*}
    &\min ||z||_1 + \sum_{i=1}^k \sum_{u \neq v} d^i(u,v) w^i_{u, v}  \text{ such that:} \\
    & z_u \ge b_u^i + \sum_{v \neq u} w^i_{v,u} - \sum_{v \neq u} w^i_{u,v} \text{ for all } u \in U, 1 \le i \le k \\
    &z_u, w^i_{u,v} \ge 0. \\
\end{align*} The variable $z_u$ represents the condition $\sum_{i=1}^k p_u^i$. The variable $w^i_{u,v}$ represents the condition $p^i_u - p^i_v \le d^i(u,v)$. 
\begin{proposition}
\label{prop:propertylp}
At an optimal solution, suppose that
\[\sum_{u \in U} b_u^i \neq \max_{j \in [k]} \left(\sum_{u \in U} b_u^j \right).\] Then there exists some $u_i \in U$ such that $p_{u_i}^i = 0$.
\end{proposition}
\begin{proof}
We prove the contrapositive. We use complementary slackness to analyze the structure of this LP. Suppose that we are at an optimal solution and $p^i_u > 0$ for all $u$ for some $i$. The corresponding condition in the dual is that $z_u = b_u^i + \sum_{v \neq u} w^i_{v,u} - \sum_{v \neq u} w^i_{u,v}$ for all $u \in U$. This means that 
\[ \sum_{u \in U} z_u = \sum_{u \in U} b_u^i + \sum_{u \in U} \sum_{v \neq u} w^i_{v,u} - \sum_{u \in U} \sum_{v \neq u} w^i_{u,v} = \sum_{u \in U} b_u^i.\]

Moreover, we know that for any $j \in [k]$, it holds that $z_u \ge b_u^j + \sum_{v \neq u} w^j_{v,u} - \sum_{v \neq u} w^j_{u,v}$ for all $u \in U$. This means that 
\[ \sum_{u \in U} z_u \ge \sum_{u \in U} b_u^j + \sum_{u \in U} \sum_{v \neq u} w^j_{v,u} - \sum_{u \in U} \sum_{v \neq u} w^j_{u,v} = \sum_{u \in U} b_u^j.\] Thus, we know that $\sum_{u \in U} b_u^i \ge  \sum_{u \in U} b_u^j$ for any $1 \le j \le k$.
\end{proof}

Now, we prove Lemma \ref{result:offlineregularrevenue}.
\begin{proof}[Proof of Lemma \ref{result:offlineregularrevenue}]
We first show this result when there is a unique advertiser $A_x$ that achieves the top bid sum. For $i \neq x$, we know that $p^i_{u_i} = 0$ for some $u_i \in U$ by Proposition \ref{prop:propertylp}. Thus, $0 \le p^i_v \le d$ for all $v \in U$ by the fairness constraint. Now, we know that the bids of advertiser $A_x$ are in the interval $[p, p+d]$ for some $p$ by the fairness constraints. Any set of bids such that $0 \le p^i_u \le d$ for $i \neq x$ and $p^x_u \in [p, p+d]$ satisfies the fairness constraints. Let's examine an individual item $u \in U$. We see that the optimal strategy is to assign $p$ mass to $A^x$ and $0$ mass to all other advertisers to start, and then distribute the remaining $1 - p$ mass by assigning $d$ mass to the top bidder, then $d$ mass to the next highest bidder, until all of the mass runs out (assigning potentially $< d$ mass to the last bidder considered).  

Let's suppose that $p \in [1-md, 1-(m-1)d]$. Let $p = 1 - (m-1)d - q$ where $0 \le q \le d$. We see that the utility achieved at $p$ is 
\[p \sum_{u \in U} b^x_u + d \sum_{i=1}^{m-1} \sum_{u \in U} ith \text{ highest bidder on u} \]
\[+ q \sum_{u \in U} mth \text{ highest bidder on u}.\] This is equal to 
\[(1- (m-1)d) \sum_{u \in U} b^x_u + d \sum_{i=1}^{m-1} ith \text{ price auction revenue} \]
\[+ q \left(-\sum_{u \in U} b^x_u + mth \text{ price auction revenue}\right).\] Thus, the maximum is achieved (not necessarily uniquely) at $q = 0$ if $\sum_{u \in U} b^x_u \ge mth \text{ price auction revenue}$ and achieved at $q = \min(d, 1 - (m-1)d)$ if $mth \text{ price auction revenue} > \sum_{u \in U} b^x_u$. 

Now, let $m$ be the maximum value such that $mth \text{ price auction revenue}$ is bigger than $\sum_{u \in U} b^x_u$ and $1 - (m-1)d \ge 0$. The above argument shows that the optimal revenue can be achieved by setting $p = (1-(m-1)d) - \min(d, 1 - (m-1)d) = \max(0, 1 - md)$. This achieves the revenue in the lemma statement (for the case where there is a unique advertiser with the top bid sum).

Now, we obtain the general case result (where multiple advertisers can attain the top bid sum). Consider all advertisers $S$ who achieve the top bid sum. If $\max_{1 \le i \le k} \sum_{u \in U} b_u^i$ is the first price auction revenue, then the statement follows by considering the mechanism that picks an arbitrary advertiser $x \in S$ and assigns every user to advertiser $x$. Otherwise, we obtain the result via a limiting argument. If $|S| > 1$, let's also pick an advertiser $x \in S$ arbitrarily. For the other advertisers $i \in S, i \neq x$, we change the bid $b^i_u$ to $b^i_u - \epsilon$ for some $u$ where advertiser $i$ is not the \textit{unique} top bid (this exists -- otherwise advertiser $i$ would be the unique advertiser with the top bid sum). We keep all of the other bids the same. Now, we can apply the result to these bids, since advertiser $x$ is the unique advertiser having the top bid sum on these new bids. It suffices to show $\epsilon \rightarrow 0$, (a) the optimal offline mechanism utility for the modified bids approaches the optimal offline mechanism utility for the original bids, and (b) the utility in the formula for the modified bids approaches the utility in the formula for the original bids (In combination, (a) and (b) imply that the utility in the formula for the original bids is the optimal offline mechanism utility for the original bids.) 

First, we show (a). Note that when bids are modified by $\pm \epsilon$, the primal LP constraints remain the same, and the objective is changed by at most $\epsilon |U|$. Thus, the optimal offline revenue changes by at most $\epsilon |U|$. 
As $\epsilon \rightarrow 0$, this goes to $0$, so we know that the optimal offline revenue is a continuous function of the bids. 

Now, we show (b). The only term in the formula that changes is the ith price revenue. We see that the ith price revenue will be dampened by at most $\epsilon \cdot |U|$, leading to at most a $\epsilon |U|$ reduction in the expression. Moreover, if $\epsilon$ is sufficiently small, then $m$ will not be affected (since we have a \textit{strict} inequality in its definition since $m > 0$). Thus, the revenue expression converges to its limit as $\epsilon \rightarrow 0$ for this sequence of bids. 
\end{proof}

We use this lemma to compute the fair value in the offline case on Example \ref{example:perfectmatching}. 
\begin{lemma}
\label{lemma:exampleofflineapproximationratio}
The fair value of the offline revenue in Example \ref{example:perfectmatching} is 
\[d + \frac{1-d}{k} + (1-d) \frac{k-1}{k} \frac{b^{low}}{b^{high}}. \]
\end{lemma}
\begin{proof}[Proof of Lemma \ref{lemma:exampleofflineapproximationratio}]
We see by Lemma \ref{result:offlineregularrevenue} that $m = 1$ since $\sum_{u \in U} b^i_u = b^{high} + (k-1)b^{low}$ while the 1st price auction revenue is $b^{high}k$ and the ith price auction revenue for $i \neq 1$ is $kb^{low}$.
\end{proof}

We see that Lemma \ref{lemma:offlineexamplerevenue} essentially follows from Lemma \ref{lemma:exampleofflineapproximationratio}. 
\begin{proof}[Proof of Lemma \ref{lemma:offlineexamplerevenue}]
We use that $\frac{b^{low}}{b^{high}} \ge \alpha^{-1}$ to obtain the desired result. 
\end{proof}

\subsubsection{Online mechanism}\label{subsubsec:online}
We now consider the online setting. 
\begin{proof}[Proof of Lemma \ref{lemma:genlowerboundbidratio}]
Suppose that all advertisers bid $1$ on the first user $u$. Suppose that the mechanism assigns the minimum probability $m$ (or tied for minimum probability) on $u$ to some advertiser $A$. Now, suppose that $A$ bids $\alpha$ on every subsequent user, and every other advertiser bids $\alpha^{-1}$ on every subsequent user. Then the first-price auction revenue is $1 + \alpha (|U| - 1)$. This mechanism's revenue is $\le 1 + (\alpha^{-1} (1-m - d) + (m + d) \alpha)(|U| - 1)$. The ratio is 
\[\le \frac{1 + (\alpha^{-1} (1-m) + m \alpha)(|U| - 1)}{1 + \alpha (|U| - 1)}.\] Since there are $k$ advertisers, we know that $m \le \frac{1}{k}$, so this is:
\[\le \frac{1 + (\alpha^{-1} (1-\frac{1}{k} - d) + (\frac{1}{k} + d) \alpha)(|U| - 1)}{1 + \alpha (|U| - 1)}.\] Let $\beta = \alpha^{-2} (1-\frac{1}{k} - d) + (\frac{1}{k} + d)$. This is equal to 
\[\beta + \frac{1 - \beta}{1 + \alpha(|U| - 1)} \] As $|U| \rightarrow \infty$, this approaches $\beta$. 
\end{proof}

\subsection{Proofs for Section \ref{subsec:gensimilar}}
First, we consider fairness of proportional allocation mechanisms.

\begin{lemma}
\label{lemma:bidratiopropalloc}
Suppose that \textbf{total variation fairness} is satisfied and we have a continuous, increasing function $g$ defining the allocation mechanism. Let $f(d)$ be the bid ratio condition. Let
\[R^{max}_g(x) := \max_{m, M \text{such that } M / m = x} \frac{g(M)}{g(m)}.\]
Then any bid ratio condition must satisfy 
\[R^{max}_g(f(d)) \le \frac{1+d}{1-d}.\] Moreover, the bid ratio condition
\[R^{max}_g(f(d)) = \frac{1+d}{1-d}.\] is sufficient. 
\end{lemma}
\begin{proof}
First, we show a sufficient bid ratio condition. 
Let's consider the difference 
\[E:= \frac{\sum_{ i \in S} C_v^i}{\sum_{j=1}^k C_v^j} -  \frac{\sum_{ i \in S} C_u^i}{\sum_{j=1}^k C_u^j},\] where  $C^i_u = g(b^i_u)$ and $C^i_v = g(b^i_v)$. Let's let $\alpha_u = \sum_{i \in S} C^i_u$, $\beta_u = \sum_{i \not\in S} C^i_u$ and $\alpha_v = \sum_{i \in S} C^i_v$, $\beta_v = \sum_{i \not\in S} C^i_v$. WLOG, assume that $E \ge 0$. Let $R_{\alpha} = \alpha_v / \alpha_u$ and let $R_\beta = \beta_u / \beta_v$. We know that $\frac{g(M)}{g(m)} \le R^{max}_g(f(d))$ by the bid ratio condition. This means that $R_{\alpha}, R_{\beta} \le  R^{max}_g(f(d))$.
We have that
\[E = 1 - \frac{\beta_v}{R_\alpha \cdot \alpha_u + \beta_v} - \frac{\alpha_u}{\alpha_u + R_\beta \cdot \beta_v}.\] Observe that this expression can be upper bounded by the case where $R_\beta$ is maximized (i.e. where $R_\beta = R^{max}_g(f(d))$) and $R_\alpha$ is maximized (i.e. where $R_\alpha = R^{max}_g(f(d))$). Our expression becomes:
\begin{align*}
    E &\le \frac{\alpha_u \cdot R^{max}_g(f(d))}{\alpha_u \cdot R^{max}_g(f(d)) + \beta_v} - \frac{\alpha_u}{\alpha_u + \beta_v \cdot R^{max}_g(f(d))} \\
     &= \frac{\alpha_u\beta_v(R^{max}_g(f(d))^2 - 1)}{(\alpha_u + \beta_v \cdot R^{max}_g(f(d)))(R^{max}_g(f(d)) \cdot \alpha_u + \beta_v)} \\
    &= \frac{\alpha_u\beta_v(R^{max}_g(f(d))^2 - 1)}{R^{max}_g(f(d)) \alpha_u^2 + R^{max}_g(f(d)) \beta_v^2 + \alpha_u\beta_v(R^{max}_g(f(d))^2 + 1)} \\
    &\le \frac{\alpha_u\beta_v(R^{max}_g(f(d))^2 - 1)}{2 R^{max}_g(f(d)) \alpha_u\beta_v + \alpha_u\beta_v(R^{max}_g(f(d))^2 + 1)} \\
    &=  \frac{R^{max}_g(f(d))^2 - 1}{2 R^{max}_g(f(d)) + R^{max}_g(f(d))^2 + 1} \\
    &=  \frac{R^{max}_g(f(d)) - 1}{R^{max}_g(f(d)) + 1}
\end{align*} 
It suffices to show that
\[\frac{R^{max}_g(f(d)) - 1}{R^{max}_g(f(d)) + 1} \le d.\] This can be solved to
\[R^{max}_g(f(d)) \le \frac{1+d}{1-d}.\]

Now, let's show a necessary bid ratio condition. Let's make $|S| = k / 2 = k'$. We suppose that on $u$, advertisers $A_1, \ldots, A_{k'}$ bid $m$ and $A_{k'+1}, \ldots, A_k$ bid $bm$, and on $v$, advertisers $A_1, \ldots, A_{k'}$ bid $bm$ and $A_{k'+1}, \ldots, A_k$ bid $m$. Now, $\alpha_u = k' g(m)$ and $\beta_u = k'g(bm)$ and $\alpha_v = k' g(bm)$ and $\beta_v = k'g(m)$. Let's set $b$, $m$ so that $R^{max}_g(f(d))$ is attained. We see that $R_{\alpha} = R^{max}_g(f(d))$ and $R_{\beta} = R^{max}_g(f(d))$. Now, we see that

\begin{align*}
   E &= 1 - \frac{\beta_v}{R_\alpha \cdot \alpha_u + \beta_v} - \frac{\alpha_u}{\alpha_u + R_\beta \cdot \beta_v}  \\
    &= \frac{\alpha_u \cdot R^{max}_g(f(d))}{\alpha_u \cdot R^{max}_g(f(d)) + \beta_v} - \frac{\alpha_u}{\alpha_u + \beta_v \cdot R^{max}_g(f(d))}.
\end{align*}

The only remaining thing to check is that the AM-GM is tight. Observe that $\alpha_u = \beta_v$ so the AM-GM is tight. 
\end{proof}

From this, Theorem \ref{thm:bidratio} follows.
\begin{proof}[Proof of Theorem \ref{thm:bidratio}]
Take $g(x) = x^l$. Observe that \[\max_{m, M \text{such that } M / m = x} \frac{g(M)}{g(m)} = \left(x\right)^l.\] The result follows.
\end{proof}

The revenue for item $u \in U$ in the proportional allocation mechanism with parameter $g(x) = x^l$ is:
\[\sum_{i=1}^k b^i_u p^i_u = \sum_{i=1}^k b^i_u \frac{(b^i_u)^l}{\sum_{j=1}^k (b^j_u)^l} = \frac{\sum_{i=1}^k (b^i_u)^{l+1}}{\sum_{i=1}^k (b^i_u)^l}. \]
Let $\vec{b}_u = [b^1_u, \ldots, b^k_u]$. Then this can be written as 
\[\frac{||\vec{b}_u||_{l+1}^{l+1}}{||\vec{b}_u||_l^l}.\]

Let's consider how this compares to a first price auction revenue for item $i$, which can be written as $||\vec{b}_u||_{\infty}$. The fair value is: 
\[\frac{||\vec{b}_u||_{l+1}^{l+1}}{||\vec{b}_u||_l^l||\vec{b}_u||_{\infty}}.\] 

Now, we compute the revenue. We begin with an analysis of $l_p$-norm relevant to the calculation. 
\begin{proposition}
\label{lpnorm}
Consider $\vec{x} \in (\mathbb{R}^{\ge 0})^n$ such that $||x||_l^l = C$. Then $||x||_{l+1}^{l+1} \ge n\left(\frac{C}{n}\right)^{\frac{l+1}{l}}$.
\end{proposition}
\begin{proof}
We prove this by induction. The base case is $n=1$, where the expression is $C^{\frac{l+1}{l}}$ as desired. Now, we do the Lagrange multipliers for $n = m$. The boundary condition is $x_i = 0$ for some number of $i$, but this just reduces to the case of a smaller $n$. We  compute the minimum for an interior point. The relevant expression is:
\[x_1^{l+1} + \ldots x_m^{l+1} - \lambda(x_1^l + \ldots x_m^l).\]
Taking a derivative of $x_i$, we obtain:
\[(l+1)x_i^l - \lambda l x_i^{l-1} = 0. \]
This can be reduced to:
\[x_l = \frac{\lambda l}{l+1}.\]
This means that all of the $x_l$ are equal, so $x_l = \left(\frac{C}{m}\right)^{l}$. Plugging this in, we obtain:
\[m\left(\frac{C}{m}\right)^{\frac{l+1}{l}}.\] This is an increasing function of $m$, so the boundary cases will not win.
\end{proof}

We prove Theorem \ref{thm:revenuebound}. 
\begin{proof}[Proof of Theorem \ref{thm:revenuebound}]
We just need to analyze $\frac{||\vec{b}_u||_{l+1}^{l+1}}{||\vec{b}_u||_l^l||\vec{b}_u||_{\infty}}$. Multiplicatively scaling by $||\vec{b}_u||_{\infty}$ leaves the expression unchanged, so we can assume WLOG that $||\vec{b}_u||_{\infty} = 1$. WLOG, let $b_u^k = 1$. Let $\vec{b'}_u = [b_1^u, \ldots, b_{k-1}^u]$. Now, the expression can be written as:
\[\frac{1 + ||\vec{b'}_u||_{l+1}^{l+1}}{1 + ||\vec{b'}_u||_l^l}. \] Now, let $C = ||\vec{b'}_u||_l^l$. For any given $C$, minimizing the expression is equivalent to minimizing $||\vec{b'}_u||_{l+1}$. By Proposition~$\ref{lpnorm}$, we claim that $||\vec{b'}_u||_{l+1}^{l+1} \ge (k-1)\left(\frac{C}{k-1}\right)^{\frac{l+1}{l}}$. Let $c =  \left(\frac{C}{k-1}\right)^{\frac{1}{l}}$. Observe that $0 \le C \le k - 1$. Then we have that our ratio is lower bounded by: 
\[\frac{1  + (k-1) \cdot c^{l+1}}{1 + (k-1) \cdot c^l} = c + \frac{1 - c}{1 + (k-1) \cdot c^l}.\]
Now, we need to minimize this expression for $0 \le c \le 1$. The derivative of this expression is equal to:
\begin{align*}
D &= \frac{(k-1)(l+1)c^l}{(k-1) c^l + 1} - \frac{(k-1)l c^{l-1} ((k-1)c^{l+1} + 1)}{((k-1) c^l + 1)^2} \\
&= \frac{(k-1)c^{l-1}}{((k-1) c^l + 1)^2}\left((l+1)c((k-1) c^l + 1) - l((k-1)c^{l+1} + 1) \right) \\
&= \frac{(k-1)c^{l-1}}{((k-1) c^l + 1)^2}\left((k-1)c^{l+1} + c(l+1) - l \right).
\end{align*}
Thus, the sign of this expression is the sign of 
\[P(c) = \left((k-1)c^{l+1} + c(l+1) - l \right).\] This expression is increasing as a function of $c$. Moreover, this expression is $k - 1 + 1 = k > 0$ at $c = 1$ and $-l$ at $c = 0$. Thus, there's exactly one root, and it occurs in the interval $c \in (0, 1)$.  Let's suppose that the root of this is $c^*$. If $P(c) < 0$, then $c < c^*$, and if $P(c) > 0$, then $c > c^*$. Then $c < c^*$, then the ratio is decreasing, and if $c > c^*$, then the ratio is increasing. 

Now, consider $c' = (k-1)^{-1/l}$. At this value, $P(c') = (k-1)(k-1)^{-(l+1/l)} + (k-1)^{-1/l} (l+1) - l = (k-1)^{-1/l} (l+2) - l$. Now, notice that $(k-1) \ge e^2 \ge \left(1 + \frac{2}{l}\right)^l$. Thus $(k-1)^{-1/l} \le \frac{1}{1 + \frac{2}{l}} = \frac{l}{l+2}$. This implies that $P(c') < 0$. Thus, we have that $c' < c^*$. 

It suffices to show that for all $c \ge c'$, the expression $c + \frac{1 - c}{1 + (k-1) \cdot c^l}$ is lower bounded by $\frac{(k-1)^{-1/l}(k-1)}{k} + \frac{1}{k}$. We observe that $c + \frac{1 - c}{1 + (k-1) \cdot c^l} \ge c + \frac{1 - c}{1 + k-1} =  \frac{c(k-1)}{k} + \frac{1}{k}$. This lower bound is an increasing function of $c$, and so we can plug in $c = c' = (k-1)^{-1/l}$ to obtain a lower bound.
\end{proof}

We prove Lemma \ref{lemma:mainpropalloclowerbound}. 

First, we show the following:
\begin{proposition}
\label{prop:mainpropalloclowerbound}
For $m < M$, let $R_b = M/m$ and let $R_g = \frac{g(M)}{g(m)}$. Let $s = \frac{1}{f^2}$ where $f$ is the fair value, and suppose that $\log_s(x)$ is a positive integer. Then we obtain that $R_g \ge R_b^{\frac{\log(k-1)}{2 \log(1/f)} - 0.5}$.  
\end{proposition}
\begin{proof}
We take bids $M, m, \ldots, m$. Let's say that we want to get a fair value of at least $f$. Note that this means:
\[f \le \frac{g(M)}{g(M) + g(m)(k-1)} + \frac{m}{M} \frac{(k-1)g(m)}{g(M) + g(m)(k-1)}. \] 
Let $R_g = g(M) / g(m)$ and let $R_b = M / m$. Then we have that
\begin{align*}
  f &\le \frac{R_g g(m)}{R_g g(m) + g(m)(k-1)} + \frac{1}{R_b} \left(\frac{(k-1)g(m)}{R_g g(m) + g(m)(k-1)}\right)  \\
  &= 1 - \frac{k-1}{R_g + k - 1} + \frac{1}{x} \left(\frac{k-1}{R_g + k-1}\right)\\
  &=  1 - \left( 1 - \frac{1}{R_b} \right)\left(\frac{k-1}{R_g + k-1}\right).
\end{align*}
To get a fair value of $f$, we need
\[\frac{k-1}{R_g + (k-1)} \le \frac{1 - f}{1 - \frac{1}{x}}. \]
This solves to 
\[R_g \ge (k-1)\left(\frac{1 - \frac{1}{x}}{1 - f} - 1 \right)= R_g \ge (k-1)\left(\frac{f - \frac{1}{R_b}}{1 - f} \right).\]

We can choose a ``seed value'' $s$ and if $\log_s(x)$ is a positive integer, then we can choose bids $m, s \cdot m$ and then $m \cdot s, m \cdot s^2$, etc. until we reach $M / s, M$.  Then the bound becomes:
\begin{align*}
    R_g &\ge \left[(k-1)\left(\frac{f - \frac{1}{s}}{1 - f} \right)\right]^{\log_s(R_b)} \\
    &= (k-1)^{\log_s(x)}\left(\frac{f - \frac{1}{s}}{1 - f} \right)^{\log_s(R_b)}  \\
    &= R_b^{\frac{\log(k-1)}{\log(s)} + \frac{\log\left(\frac{f - \frac{1}{s}}{1 - f} \right)}{\log(s)}}.
\end{align*}
Let's take $s = \frac{1}{f^c}$. Then we obtain $\frac{\log(k-1)}{c \log(1/f)} + \frac{\log(\frac{f-f^c}{1-f})}{c \log(1/f)}$. If we set $c = 2$ then we obtain $\frac{\log(k-1)}{2 \log(1/f)} - 0.5$.
\end{proof}

We use this bound to prove Lemma \ref{lemma:mainpropalloclowerbound}.

\begin{proof}[Proof of Lemma \ref{lemma:mainpropalloclowerbound}]
By Proposition \ref{prop:mainpropalloclowerbound}, we have that
\[R_g\ge R_b^{\frac{\log(k-1)}{2 \log(1/f)} - 0.5}\] when $\log_{1/f^2}(R_b)$ is a positive integer. We extend this result to when $\log_{1/f^2}(R_b) = 1 /n$ for $n \in \mathbb{N}$. More specifically, we show that 
\[R_g^{max}(x) \ge x^{\frac{\log(k-1)}{2 \log(1/f)} - 0.5}\] where $R_g^{max}(x) = \max_{m, M \text{ such that } M / m = x} \frac{g(M)}{g(m)}$.  

Let's take $m, m \cdot \left(\frac{1}{f^2}\right)^{1/n}, m \cdot \left(\frac{1}{f^2}\right)^{2/n}, \ldots, m \cdot \left(\frac{1}{f^2}\right) =: M$. We see that 
\[\left(\frac{M}{m}\right)^{\frac{\log(k-1)}{2 \log(1/f)} - 0.5} \le \frac{g(M)}{g(m)} = \prod_{i=1}^{n} \frac{g(m \cdot \left(\frac{1}{f^2}\right)^{i/n})}{g(m \cdot \left(\frac{1}{f^2}\right)^{(i-1)/n})}.\] This implies that there exists $i$ such that 
\[R_g^{max}(\left(\frac{M}{m}\right)^{1/n}) \ge \frac{g(m \cdot \left(\frac{1}{f^2}\right)^{i/n})}{g(m \cdot \left(\frac{1}{f^2}\right)^{(i-1)/n})} \ge \left(\left(\frac{M}{m}\right)^{1/n}\right)^{\frac{\log(k-1)}{2 \log(1/f)} - 0.5}.\] This provides the desired statement. Now, we plug this into Lemma \ref{lemma:bidratiopropalloc}. We see that if $f(d)$ is of the form given in the lemma statement, then we know that 
\[\left(f(d)\right)^{\frac{\log(k-1)}{2 \log(1/f)} - 0.5} \le \frac{1+d}{1-d}.\] Solving gives us that:
\[f(d) \le \left(\frac{1+d}{1-d}\right)^{1/l} \] where $l = \frac{\log(k-1)}{2 \log(1/f)} - 0.5$. 
\end{proof}

\section{Proofs for Section 5}


We now prove Proposition \ref{prop:lowerboundofflinedifferent}.
\begin{proof}[Proof of Proposition \ref{prop:lowerboundofflinedifferent}]
We actually show the result for a generalization of Example \ref{example:wantedperson} where advertiser $i$ places bids of $b^{high}$ on $u_i$ and $u_{c+1}$ and $b^{low}$ on everyone else, and where $d^i(u_{c+1}, u_i) = d$. Observe that the first-price auction revenue is at least $b^{high}(c+1)$. 

If advertiser $i$ receives the ``jack-of-all-trades'' with probability $p^i_{u_{c+1}}$, then we know that that $\sum_{i=1}^c p^i_{u_{c+1}} = 1$. Moreover, the total utility becomes bounded by $b^{high}(\min(1,p^1_{u_{c+1}}+ d)) + b^{high}(\min(1,p^2_{u_{c+1}} + d)) + \ldots + b^{high}(\min(1,p^k_{u_{c+1}} + d)) + b^{high} \cdot (p^1_{u_{c+1}} + \ldots + p^n_{u_{c+1}}) + b^{low} (\text{ leftovers }) \le 2b^{high} + b^{high} d c + b^{low} \cdot c$.

Thus, the fair value is at most $\frac{2}{c+1} + d + \frac{b^{low}}{b^{high}} = \frac{2}{c+1} + d + \frac{1}{f(1)}$
\end{proof}

We prove Theorem \ref{thm:upperbounduserspecified}.
\begin{proof}[Proof of Theorem \ref{thm:upperbounduserspecified}]
First, observe that the total mass sums to at most $1$ since $R + R(\beta^{-1} + \ldots + \beta^{-C}) \le 1$. Now, we show that the fairness properties are satisfied. Let's suppose that a user specifies $D$ categories. Then there is at least a mass of $R (\beta^{-1} + \ldots + \beta^{-D})$ on these categories. So, we must show that the mass of these categories on any other user is bounded by $R + R(\beta^{-1} + \ldots + \beta^{-D +1})$. If another user specifies $C' < D$ categories, then the mass on these categories is at most $R + R(\beta^{-1} + \ldots + \beta^{-C'}) \le R(1 + \beta^{-1} + \ldots + \beta^{-D +1})$ as desired. If another user specifies $C' \ge D$ categories, then the mass is at most $R + R(\beta^{-1} + \ldots + \beta^{-C'}) \frac{D-1}{C'} \le R + R(\beta^{-1} + \ldots + \beta^{-D +1})$ as desired.  
\end{proof}

We prove Lemma \ref{lemma:lowerbounduserspecified}.
\begin{proof}[Proof of Lemma \ref{lemma:lowerbounduserspecified}]
Suppose that an online allocation mechanism achieves an fair value of $>R$ and $\beta$ inter-category envy-freeness.

We will specify a sequence of user types $t_0, t_1, \ldots, t_n, \ldots, t_{C}$. A user of type $t_0$ will specify a set $S_{t_0}$ of $1$ category and a user of type $t_n$ for $n \ge 1$  will specify a set $S_{t_n}$ of $n$ categories. For users of type $t_n$, on all of the categories, except for a special category $C_n \not\in S_{t_n}$, the bid is $0$ on the user, while the bid on $C_n$ is $1$.

We will adaptively construct the sets $S_{t_n}$. The category $C_n$ will be selected from the set of categories not in $S_{t_n}$. (Recall that $C$ is less than the total number of categories, so there is always at least one such category.)

The adversary will perform the following high-level strategy. It will iterate through users of types $t_0$, then type $t_1$, etc. It will continue to specify users of type $t_n$ while the probability placed on $C_n$ is less than $R$. If $n < C$, if the probability is at least $R$, then the adversary switches to $t_{n+1}$. If the process gets stuck at some $t_n$ for $n < C-1$, then the fair value will necessarily be at most $R$. Thus, since the fair value is greater than $R$, we know that the process will reach $t_C$ at some finite time step. 
 
Now, we show how to construct the sets $S_{t_n}$ so that for $n \ge 1$, there is at least a probability of $R\beta^{-1} + R \beta^{-2} + \ldots + R \beta^{-n}$ must be placed on $S_{t_n}$ for users of type $t_n$. We construct the sets inductively. The first set $S_{t_0}$ can be any category. For $S_{t_1}$, we pick $C_0$. Since we switched to type $t_1$ users, there was at least a mass of $R$ placed on $C_0$ in the last type $t_0$ user. Thus, there must be at least a $R \beta^{-1}$ mass on $S_{t_1}$ by users of type $t_1$ as desired. For $n \ge 2$, for set $S_{t_{n+1}}$, we use the fact that there is at least a $R\beta^{-1} + R \beta^{-2} + \ldots + R \beta^{-n}$ mass on $S_{t_{n}}$. We let $S_{t_{n+1}}$ be $S_{t_n}$ coupled with $C_n$. The mass on $C_n$ on the last user of type $t_n$, since we switched to $t_{n+1}$, is at least $R$. Thus, the total mass on $C_n$ and $S_{t_n}$ on this type $t_n$ user is $R + R\beta^{-1} + R \beta^{-2} + \ldots + R \beta^{-n}$. Thus, there must be at least a $R\beta^{-1} + R\beta^{-2} + R \beta^{-3} + \ldots + R \beta^{-{n+1}}$ on $S_{t_{n+1}}$ on type $t_{n+1}$ users as desired. 

In order to have a fair value of $R$, there must be some type $t_C$ user that has a mass of $R$ on $C_{C}$. Thus, the total mass on advertisers in $S_{t_C}$ and $C_C$ is at least $R + R \beta^{-1} + \ldots + R\beta^{-C}$. Thus, we have that $R + R \beta^{-1} + \ldots + R\beta^{-C} \le 1$, so $R \le \frac{1}{1 + \beta^{-1} + \ldots + \beta^{-C}}$ as desired. 
\end{proof}

We prove Theorem \ref{thm:userdirectedpositiveresult}.
\begin{proof}[Proof of Theorem \ref{thm:userdirectedpositiveresult}]
This follows from the fact that the revenue in Mechanism \ref{mech:firstpricevariant} is $\sum_{u \in U} \max_{i \in S_u} b^i_u$ and the probability of selecting a category in $S_u$ on $u$ is $1$.
\end{proof}

\section{Proofs for Section 6}

\begin{proof}[Proof of Theorem \ref{thm:composition1}, Proof of Theorem \ref{thm:composition2}]
Fairness follows from the fairness of Mechanism \ref{mech:userspecified} and Mechanism \ref{mech:firstpricevariant}, as well as the fairness of Mechanism \ref{mech:propalloc} in Theorem \ref{thm:bidratio} and since 
Mechanism \ref{mech:propalloc} uses the full probability mass on each user. The fair value easily follows from the fair value of Mechanism \ref{mech:propalloc} in Theorem \ref{thm:revenuebound} and the fair value of Mechanism \ref{mech:firstpricevariant} in Theorem \ref{thm:userdirectedpositiveresult}
 and Mechanism \ref{mech:userspecified} in Theorem \ref{thm:upperbounduserspecified}.
 \end{proof}

\section{Mechanisms for uniform metrics}\label{sec:uniformmechanisms}

We consider the following mechanism for the uniform metric setting and show upper bound nearly matches the lower bounds in Section \ref{subsec:uniform}. 
\begin{mech}[Shifted Mechanism]
\label{mechanism:uniformshifted}
The mechanism assigns a top bidder for the first user $1$ and all other bidders $0$. On future users, if the advertiser does not have the top bid, then the mechanism assigns $1-d$ to that advertiser and $d$ to the advertiser with the top bid, and $0$ to other advertisers.
\end{mech} 
\noindent Observe that the fair value achieved by this mechanism is at least $(1 - d)\alpha^{-2} + d$. This is very close to (in fact within an additive error of $\frac{1}{k}$ of) the upper bound in Lemma \ref{lemma:genlowerboundbidratio}. This demonstrates that the upper bound in Lemma \ref{lemma:genlowerboundbidratio} essentially cannot be tightened using uniform metrics. 

Since this mechanism requires use of the metric and identities, we also consider other mechanisms, though the fair value of these mechanisms is worse.
The next mechanism is metric-oblivious and is fair for non-uniform metrics but is highly asymmetric based on bids on the first user. It does not provide good revenue guarantees when $d \rightarrow 1$ where the bid ratio goes to $\infty$. 
\begin{mech}[Top Bidder Mechanism]
\label{mech:uniformtop}
For item $u \in U$ and $1 \le i \le k$, the mechanism assigns a top bidder on the first user $p_u^{i} = 1$ for all bids.
\end{mech}
\noindent It is straightforward to see that the fair value is $\alpha^{-2}$. 

The next mechanism is a modification of the previous mechanism  that achieves good revenue guarantees, but that depends on $d$. It continues to have a strong asymmetry based on bids on the first user. 
Now, we consider a symmetric mechanism that also depends on $d$. However, it does not provide reasonable revenue guarantees.
\begin{mech}[Even Mechanism]
\label{mech:uniformeven}
For item $u \in U$ and $1 \le i \le k$, the mechanism assigns a top bidder $p_u^{i} = \frac{1}{k} + \frac{d(k-1)}{k}$ and all other bidders $\frac{1}{k} - \frac{d}{k}$.
\end{mech}
\noindent It is straightforward to see that the fair value is $\frac{1}{k} + \frac{d(k-1)}{k}$.

The next mechanism is a modification of the previous mechanism that  breaks some of the symmetry of the previous mechanism by performing weeding out of advertisers based on bids on the first user. it provides much better revenue guarantees. 
\begin{mech}[Improved Even Mechanism]
\label{mech:uniformimprovedeven}
For item $u \in U$ and $1 \le i \le k$, the mechanism assigns a probably of $0$ to all advertisers that aren't within $\alpha^{-2}$ of the top advertiser on the first user $0$. The mechanism assigns a top bidder $p_u^{i} = \frac{1}{k} + \frac{d(l-1)}{l}$ and all other bidders $\frac{1}{l} - \frac{d}{l}$ to the $l$ remaining advertisers. 
\end{mech}
\noindent It is straightforward to see that the fair value is at least $\frac{1}{k} + \frac{d(k-1)}{k} + \alpha^{-4} \left(\frac{k-1}{k} - \frac{d(k-1)}{k} \right)$.

\section{Relaxations for the identical metric case}\label{sec:identicalrelaxations}
One possible relaxation is to consider $(1 - \delta)$-relaxed total variation fairness, a Lipschitz relaxation of total variation fairness. This definition essentially disregards distances $> 1 - \delta$ and scales the other distances out.  \begin{definition}[Relaxed Total Variation Fairness]
A central body mechanism satisfies $(1-\delta)$-relaxed-\textbf{total variation fairness} for a metric $d$ and constant $\delta > 0$ if $|\sum_{i \in S} p^i_u - \sum_{i \in S} p^i_v| \le (1-\delta)^{-1} d(u,v)$ for every $1 \le i \le k$, $u,v \in U$, and \textit{all} subsets $S$ of advertisers.  
\end{definition}
We can easily obtain the following by applying the mechanism to $d(u,v) \cdot (1 - \delta)$ and ignoring distances bigger than $1 - \delta$. 
\begin{proposition}
\label{prop:relaxedgeneralizedimprovedbidratio}
With the bid ratio condition 
\[f(d(u,v)) = \left(\frac{1+(1-\delta)^{-1}d(u,v)}{1-(1-\delta)^{-1}d(u,v)}\right)^{1/l}\] for $d(u,v) \le 1 - \delta$ for some parameter $g(x) = x^l$ for $l \ge 1$, Mechanism \ref{mech:propalloc} satisfies $(1-\delta)$-relaxed total variation fairness.
\end{proposition}
This bid ratio has the nice property that as $d \rightarrow 1-\delta$, the bid ratio condition goes to $\infty$. Moreover, the bid ratio is weaker than the bid ratio condition in Section \ref{subsec:gensimilar}.

\section{Metric Relaxations of Inter-Category Envy-freeness}\label{sec:differentmetricslowerbounds}
We consider a weaker notion of fairness based on additively relaxing based on the sum of fairness metrics in the fairness constraints.
\begin{definition}[Inter-Metric Envy-Freeness]
A mechanism satisfies $\beta$-\textbf{inter-metric envy-freeness} if for all $u \in U$, it is true that: $\sum_{i \in S_u} p_v^i \le \beta \left(\sum_{i \in S_u} p_u^i + \sum_{i \in S_u} d^i(u,v)\right)$. 
\end{definition}
We prove lower bounds by considering the following example: 
\begin{example}
\label{example:onlinerelaxedOR}
Suppose for all $u, v \in U$, we have that $d^i(u,v) \le d_{small}$ for $1 \le i \le j$ and $d^i(u,v) \ge d_{big}$ for $j+1 \le i \le c$. On all users, the bids on $C_1, \ldots, C_j$ are $b$, for some constant $b > 1$.  Suppose that we can partition $U$ into $U_1 \cup U_2$ where users in $U_1$ are type 1 and users in $U_2$ are type 2. For $u \in U_1$ we have that the bids on $C_{j+1}, \ldots, C_c$ are $M$ for $M > b$ and for $u \in U_2$, we have that the bids on $C_{j+1}, \ldots, C_c$ are $1$. Let's suppose that all type $1$ users specify $C_1, \ldots, C_j$ for its selected categories\footnote{We don't even need type 2 users to specify fairness constraints.}.
\end{example}
The idea if $b >> 1$ and $M >> b$, then on type 1 users, the bids from $C_1, \ldots, C_j$ are significantly worse than the bids from $C_{j+1}, \ldots, C_c$, while on type 2 users, the bids from $C_1, \ldots, C_j$ are significantly better than the bids from $C_{j+1}, \ldots, C_c$. Since the type 1 users requested fairness on $C_1, \ldots, C_j$, it is not possible to simultaneously place a high mass on $C_{j+1}, \ldots, C_c$ for type 1 users and $C_1, \ldots, C_j$ for type 2 users. We specifically show the following lower bounds:
\begin{proposition}
\label{prop:lowerbounduserconstraints}
Consider online mechanisms for different metrics that achieve $\beta$ inter-metric envy-freeness. 
\begin{enumerate}
    \item If the bid ratio condition is $\infty$ at $1$ and distances of $0$ are permitted, then any such mechanism has a fair value of $R \le \frac{\beta}{\beta + 1}$.
    \item If the bid ratio condition is $\infty$ at $1$, users are only permitted to specify $\le C$ categories, and $d(u,v) \ge d_{small}$ for $u \neq v$, then any such mechanism has a fair value of $R \le  \frac{\beta}{\beta + 1} \left(1 + C \cdot d_{small}\right)$. 
    \item Suppose that $f(1) < \infty$. Then any such mechanism has a fair value of $R \le \frac{\beta}{\beta + 1} + \frac{1}{\sqrt{f(1)}(\beta + 1)}$. 
\end{enumerate}
\end{proposition}
\begin{proof}[Proof of Proposition \ref{prop:lowerbounduserconstraints}]
It suffices to show the following: if the metric has minimum distance $d_{small}$, each category has $\le C$ advertisers, and $m$ is the minimum-to-maximum bid ratio condition, then any online mechanism that achieves $\beta$-inter-metric envy-freeness online mechanism has a fair value $R \le \frac{\beta}{\beta + 1}(1 + C \cdot d_{small}) + \frac{(1 - C \cdot d_{small} \beta) \sqrt{m}}{\beta+1}$.

The proof relies on Example \ref{example:onlinerelaxedOR}. Let $\delta$ be some value and suppose there are $\ge b / \delta$ users. 

Let's consider a type $1$ user $u$ and type $2$ user $v$. Let $p_1 := \sum_{i=1}^j p^i_u$ and $p_2 := \sum_{i=1}^j p^i_v$. Suppose that we have a fairness condition that says that $p_2 \le \beta(p_1 + D)$. 

On $u$, the fair value is $1 - p_1 + p_1 \frac{b}{M} = 1 - p_1 \left(1 - \frac{b}{M}\right)$. On $v$, the fair value is $p_2 + \frac{1}{b}(1 - p_2) = \frac{1}{b} + p_2 \left( 1 - \frac{1}{b}\right)$. Now, our fairness condition says that $p_2 \le \beta (p_1 + D)$. Thus we know that $\frac{1}{b} + p_2 \left( 1 - \frac{1}{b}\right) \le \frac{1}{b} + \beta p_1 \left( 1 - \frac{1}{b}\right) + \beta D \left(1 - \frac{1}{b}\right)$.

To maximize the fair value, we need to set
$1 - p_1 \left(1 - \frac{b}{M}\right) = \frac{1}{b} + \beta p_1 \left( 1 - \frac{1}{b}\right) + \beta D \left(1 - \frac{1}{b}\right)$. This solves to 
\[p_1 = \frac{(1 - \frac{1}{b})(1-D \beta)}{\beta(1 - \frac{1}{b}) + 1 -  \frac{b}{M}}.\] Now we use the fact that $M = b^2$ to obtain:
\[p_1 = \frac{1-D \beta}{\beta + 1}.\] This implies that the fair value is at most: \[1 - p_1 \left(1 - \frac{b}{M}\right) = 1 - \left(1 - \frac{1}{b}\right) \frac{1-D \beta}{\beta + 1} = \frac{\beta}{\beta+1}(1 + D) + \frac{1-D\beta}{(\beta + 1)b}.\] 

What is the specific order of user arrivals? We first send in a type $1$ user. We continue to send in type $1$ users while the total probability assigned to $C_1, \ldots, C_j$ is $\le \frac{1-D \beta}{\beta + 1}$. As shown above, the fair value is at least $ \frac{\beta}{\beta+1}(1 + D) + \frac{1-D\beta}{(\beta + 1)b}$. If the total probability exceeds $\frac{1-D \beta}{\beta + 1}$ at any point, then we switch to sending in type $2$ users. Now, the fair value on type 2 users is at least $\le \frac{1-D \beta}{\beta + 1}$ as well.

Thus, the fair value is at least $r = \frac{1-D \beta}{\beta + 1}$ except on potentially one user where the switch to type 2 users occurs. The total fair value can be upper bounded by $\frac{M + rQ}{m+Q}$. As long as $Q \ge M / \delta$, we can bound this by $r + \delta$. We can phrase this as long as there are $\ge b / \delta$ users. 

Let's set $\delta \rightarrow 0$ and observe that $b \le \frac{1}{\sqrt{m}}$ to obtain the desired expression. 
\end{proof}

\section{Multiple-Task Fairness Against Relaxed Benchmark}\label{sec:multipletaskrelaxed}

We now show that with the relaxed fair value based on user preferences, it is still not possible to recover multiple-task fairness-style guarantees. The fairness notion that we consider is a inter-metric envy-freeness variant of version of multiple-task fairness. 
\begin{definition}[Inter-Metric Multiple-Task Envy-Freeness]
A mechanism satisfies $\beta$-\textbf{inter-metric multiple-task envy-freeness} with respect to some function $f$ if $p^i_v \le \beta(p^i_u + d^i(u,v))$ for all $i \in S_u$ and $u \in U$. 
\end{definition}

First, we show that if we don't place restrictions on the subsets $S_u$, then we can't achieve fairness w.r.t \text{any} reasonable combination of the metrics. (Here, observe that the guarantees don't necessarily get stronger as $C$ increases because the fair value definition also changes. In fact in the examples we consider, the guarantees get weaker in a sense.) The bound makes it so that some users specify a set of categories with low bids, while other users swap out of one of the categories for a category with a high bid.
\begin{proposition}
\label{prop:lowerboundsubsetsoverlapping}
Even with a maximum-to-minimum bid ratio condition of $1$ on the advertisers at all distances (i.e. each category always has the same bids on all users), a mechanism that receives sets $S_u$ of size $C$ and satisfies inter-metric multiple-task envy-freeness has a fair value of at most:
\[R \le \frac{\beta}{\beta + \frac{C-1}{C}} + \frac{d_{small}(C-1)}{\beta + \frac{C-1}{C}},\]
if the metric satisfies $d \ge d_{small}$. 
\end{proposition}
\begin{proof}[Proof of Proposition \ref{prop:lowerboundsubsetsoverlapping}]
Suppose that we have a fair mechanism achieving a fair value of greater than $R$. Suppose that we have categories $C_1, \ldots, C_{c}$. Let's suppose the distance metric is $d$ everywhere. Let's suppose that $C_1, \ldots, C_j$ have bids of $1$ on all users and $C_{j+1}, \ldots, C_c$ have a bid of $B > 1$ on all users.  Let's give the mechanism a user $u$ that specifies $C_1, \ldots, C_j$. Suppose the mechanism places a total probability of $p$ on $u$. If $p < R$, we repeat identical users until the mechanism assigns more mass on this user. Thus eventually $p > R$. Now, there is some set of $C-1$ advertisers that have a total of $\ge \frac{p(C-1)}{C}$ mass. Let's give the mechanism a user that specifies these categories and $C_{j+1}$. The mechanism has to continue to place a mass of $\frac{1}{\beta} \left(\frac{p(C-1)}{C} - d(C-1)\right)$ on these advertisers. Now, the mechanism can only put $1 - \frac{1}{\beta} \left(\frac{p(C-1)}{C} - d(C-1)\right) < 1 - \frac{1}{\beta} \left(\frac{R(C-1)}{C} - d(C-1)\right)$ mass on $B$. Since we can make $B$ arbitrarily large, it must be true that:
 \[ R \le 1 - \frac{1}{\beta} \left(\frac{R(C-1)}{C} - d(C-1)\right).\]
 This gives us:
 \[\beta - \beta R \le  \frac{R(C-1)}{C} - d(C-1).\] We can further solve to obtain the desired bound.
 \end{proof}

Now, let's try to lower our expectations and allow the user to choose from a collection of prescribed sets $S_1, \ldots, S_l$ that partition the categories so that these sets are non-intersecting. Now, if advertiser bid the same on all users, the problem is simple (run a first-price auction in each $S_i$). Without these extremely restrictive conditions, it turns out to still be impossible to achieve inter-metric multiple-task envy-freeness with a good fair value. These bounds use Example \ref{example:onlinerelaxedOR}.  
\begin{proposition}
\label{prop:lowerboundusermultipletask}
Consider online mechanisms for different metrics that achieve $\beta$-inter-metric multiple-task envy-freeness. Let's suppose that users can choose from prescribed category choices $S_1, \ldots, S_l$,. 
\begin{enumerate}
    \item If the bid ratio condition is $\infty$ at $1$ and distances of $0$ are permitted, then any such mechanism has a fair value of $R \le \frac{\beta}{\beta + 1}$.
    \item If $f(1) < \infty$, then any such mechanism has a fair value of $R \le \frac{\beta}{\beta + 1} + \frac{1}{\sqrt{f(1)}(\beta + 1)}$. 
\end{enumerate}
\end{proposition}
\begin{proof}[Proof of Proposition \ref{prop:lowerboundusermultipletask}]
We use Example \ref{example:onlinerelaxedOR}. Let 
\[S_1 = \left\{C_1, C_{j+1}, C_{j+2}, \ldots, C_{c}\right\}\] and suppose all users pick $S_1$. We see that in this example, the utility on $\sum_{u \in U} \max_{i \in S_1} b^i_u$ matches the first-price utility, so the fair value and relaxed value will be identical. Now, we use the fact that if a mechanism satisfies $\beta$-inter-metric multiple-task envy-freeness, then it satisfies $\beta$-inter-metric envy-freeness on $\left\{C_1\right\}$. Thus, by setting $j = 1$, the lower bounds from the proof of Proposition \ref{prop:lowerbounduserconstraints} on this example apply in this setting. 
\end{proof}

\section{Position Auctions}\label{sec:positionauctions}
Roughly speaking, a position auction only uses the ordering of the bids, without any information about the identities of the advertisers or the bid values. The first-price auction, which produces the optimal utility in the absence of fairness, is a position auction that selects the highest bidder with probability $1$. We show that with multiple-task fairness constraints, it is impossible for a position auction to achieve a competitive revenue. This example demonstrates why we must give the platform greater information about the bids in order to simultaneously achieve fairness and a competitive fair value. 

In our upper bound, we specifically consider the setting where the platform is only given access to the ordering of current bids (and not the identities of advertisers or the values of the current bids) on the current user. \textit{After} selecting a fractional allocation for this user, the platform is giving access to the values of the bids and the identities of the advertisers on those bids. Intuitively, not knowing the identity of the highest bidder or values of the other bids makes it difficult to place a high probability on the highest bid while satisfying fairness constraints. We show that with \textit{any} non-trivial bid ratio condition, it is impossible for a position auction with fractional allocations to simultaneously achieve multiple-task fairness and a high fair value. 
\begin{lemma}
\label{lemma:lowerboundrestricteduniform}
If the metrics are identical and uniform (i.e. $d^i(u,v) = d$ for all $u, v \in U$) and with \textbf{any} bid ratio condition $> 1$ for $d > 0$, an online mechanism that only sees the ordering of the bids and achieves multiple-task fairness has a fair value of $\le \frac{1}{k} + 2d$.
\end{lemma}
 
\begin{proof}[Proof of Lemma \ref{lemma:lowerboundrestricteduniform}]
First, we show an online mechanism that only sees the order of the bids must place at most a $\frac{1}{k} + d$ probability on the top bidder, if there is a unique top bid. We consider bid sequences with a unique top bid and all other bids equal. 

Let $p_1$ be the probability on the top bid on the first user. First, we show that fairness alone tells us that $p_1 \le \frac{1}{k} + d$. Let's say that the first user has bids $1, 1-\delta, 1 - \delta, \ldots, 1- \delta$. Now, the second user can (approximately) have bids $\alpha, \alpha^{-1}, \ldots, \alpha^{-1}$. Assuming that $\alpha > 1$, this means that any permutation is possible and we can't differentiate between permutations. Let's say that we put a probability of $S$ on the top advertiser on the next user. We can choose an assignment where the top advertiser from the first user is the top advertiser from the second user, so $S \ge p_1 - d$. We can also choose an assignment where the top advertiser on the first user is assigned to the minimum probability on the second user. This means that $\frac{1-S}{k-1} \ge p_1-d$. (Solving, we obtain $S \le 1 - (k-1)(p_1-d)$.) This implies that $kp_1 - dk \le S + (k-1) \frac{1-S}{k-1} = 1$, so $p_1 \le \frac{1}{k} + d$. 

Let's suppose the first user actually has bids $1, 0, 0, \ldots, 0$. Now, the top bidder can only receive $\frac{1}{k} + 2d$ on any subsequent bid, so the fair value is at most $\frac{1}{k} + 2d$.  
\end{proof}
\noindent We see that in the limit as $k \rightarrow \infty$ and $\epsilon \rightarrow 0$, this condition disallows $d(u,v) \le \frac{1}{2}$. This condition is far too strong to permit the necessary level of expression for the fairness metric.

\section{Convexity of the bid ratio constraint}\label{sec:convexitybidratio}
The family of constraints described in Section \ref{subsec:bidratio} has the following super-multiplicative-like property at all points: if $0 < d_1, d_2 < 1$, then $f(d_1 + d_2) > f(d_1) f(d_2)$. This has the following peculiar consequence. Suppose that $u_1$ and $u_2$ satisfy that $d(u_1, u_2) = d$ where $d > 0$. In this case, the bid ratio condition is $\frac{1}{f(d)} \le \frac{b_{u_1}}{b_{u_2}} \le f(d)$. Suppose that $u_3$ is on a line in between $u_1$ and $u_2$ (i.e. that $d(u_1, u_3) + d(u_3, u_2) = d(u_1, u_2)$, where $u_3$ is not at either endpoint (i.e. that $d(u_3, u_1), d(u_3, u_2) \neq 0$. Including the user $u_3$ in between $u_1$ and $u_2$ necessarily \textit{strengthens} the bid ratio condition between $u_1$ and $u_2$ since $\frac{b_{u_1}}{b_{u_2}} = \frac{b_{u_1}}{b_{u_3}} \cdot \frac{b_{u_3}}{b_{u_2}}$ which is upper bounded by $f(d(u_3, u_1))f(d(u_3, u_2)) < f(d(u_1, u_2))$ and lower bounded by $\frac{1}{f(d(u_3, u_1))f(d(u_3, u_2))} > \frac{1}{f(d(u_1, u_2))}$. This means that the users cannot be specified in an online manner to the advertisers: the advertiser may bid on users in a way that satisfies the bid ratio conditions on $u$ and $v$ at the current time step but violates it on a future time step.

If a bid ratio condition $f$ does not have this super-multiplicative-like property at any points (i.e. that $f(d_1 + d_2) \le f(d_1) f(d_2)$ for all $d_1, d_2 \ge 0$), then this means that the bid ratio condition is concave or linear. We now briefly consider concave bid ratio constraints with a finite bound on $f(1)$ and place a restrictive upper bound on the fair value. If $f(1)$ is permitted to be a large finite number, it is not possible to have a concave bid ratio condition along with competitive fair value in an offline mechanism that achieves multiple-task fairness. 
\begin{lemma}
\label{lemma:bidratioconcavelowerbound}
Suppose that the bid ratio condition satisfies $f(0) = 1$ and $f(1) = h$ and is concave (or linear). Then, any offline mechanism that satisfies multiple-task fairness on a general metric has a fair value $\le \frac{1}{k} + \frac{2}{\sqrt{h-1}}$.  
\end{lemma}
\begin{proof}
Any concave function $f$ will satisfy $f(d) \ge 1 + d(h-1)$. Moreover, by Lemma \ref{lemma:offlineexamplerevenue}, we know that the fair value is at most $R \le \frac{1}{k} + \frac{1}{1 + d(h-1)} + d$. At $d = \frac{1}{\sqrt{h-1}}$, we see that the fair value is at most $\frac{1}{k} + \frac{1}{\sqrt{h-1}} + \frac{1}{1 + \sqrt{h-1}} \le \frac{1}{k} + \frac{2}{\sqrt{h-1}}$ as desired. 
\end{proof}
\noindent When $h$ is large, this fair value is small. This demonstrates that to achieve a high fair value and multiple-task fairness, we cannot restrict to concave (or linear) bid ratio functions.

\end{document}